\documentclass[aps,prd,nofootinbib,showkeys,floatfix,twocolumn]{revtex4-1}
\usepackage{bbold}
\usepackage{amsmath}
\usepackage{amssymb}
\usepackage{graphicx}
\usepackage{xspace}
\usepackage{bbm} 
\usepackage{color}
\usepackage[utf8]{inputenc}
\usepackage{cancel}
\usepackage{slashed}
\usepackage{soul}
\usepackage{units}
\usepackage{enumitem}

\usepackage{tikz} 
\usetikzlibrary{chains,shapes,fit,calc}
\usetikzlibrary{positioning,arrows}

\usetikzlibrary{patterns,decorations.pathreplacing}
\usetikzlibrary{decorations.pathmorphing}
\usetikzlibrary{decorations.markings}
\usetikzlibrary{arrows,shapes}
\usetikzlibrary{shapes.misc}
\tikzset{
    gluon/.style={decorate, draw=black,
        decoration={coil,amplitude=4pt, segment length=4pt,aspect=0.7}} 
}
\tikzset{
    photon/.style={decorate, decoration={snake}},
}

\newcommand\SARAH{{\tt SARAH}\xspace}

\newcommand\SPheno{{\tt SPheno}\xspace}

\newcommand\micro{{\tt MicrOMEGAs}\xspace}
\newcommand\higgsb{{\tt HiggsBounds}\xspace}

\newcommand{\Lag}{\mathcal{L}}
\newcommand{\Wsuper}{\mathcal{W}}
\newcommand{\order}[1]{\mathcal{O}(#1)}
\newcommand{\msusy}{M_{\tiny SUSY}}
\newcommand{\vev}{\textit{v}_\textit{\tiny SM}}

\newcommand{\vs}{\textit{v}_\textit{\tiny S}}

\newcommand{\ccdot}{\!\cdot\!}

\newcommand{\tb}{t_{\beta}}

\def\Sr{S_r}
\def\Si{S_i}
\def\tW{\tilde{W}}
\def\tG{\tilde{g}}
\def\tB{\tilde{B}}
\def\tHu{{\tilde{H}_u}}
\def\tHd{{\tilde{H}_d}}
\def\tS{\tilde{S}}

\definecolor{fsblue}{rgb}{0.,.0,1.}

\usepackage[pdftex,colorlinks,linkcolor=cyan,citecolor=blue,urlcolor=cyan]{hyperref} 
\hypersetup{linktocpage}
\usepackage[capitalise, english]{cleveref}
\crefname{section}{Sec.}{Secs.}
\crefname{table}{Tab.}{Tabs.}

\graphicspath{{figs/}}
\makeatletter
\def\input@path{{chapters/}}
\makeatother

\begin{document}
\vspace{1cm}

\title{\LARGE The Singlet Extended Standard Model \\ in the Context of Split Supersymmetry}

\hfill \parbox{5cm}{\vspace{ -1cm } \flushright P3H-19-018 KA-TP-14-2019}

\newcommand{\AddrKAITP}{
Institute for Theoretical Physics (ITP), 
Karlsruhe Institute of Technology, \\
Wolfgang-Gaede-Stra{\ss}e 1, D-76131 Karlsruhe, Germany}

\newcommand{\AddrKAIKP}{
Institute for Nuclear Physics (IKP), Karlsruhe Institute of Technology,\\
Hermann-von-Helmholtz-Platz 1, D-76344 Eggenstein-Leopoldshafen, Germany}

\author{Martin Gabelmann} \email{martin.gabelmann@kit.edu} 
\affiliation{\AddrKAITP}

\author{M. Margarete M\"uhlleitner} \email{milada.muehlleitner@kit.edu} 
\affiliation{\AddrKAITP}

\author{Florian Staub}\email{florian.staub@kit.edu}
\affiliation{\AddrKAITP}\affiliation{\AddrKAIKP}%

\begin{abstract}
    We consider a low-energy effective theory of the next-to-minimal
    supersymmetric Standard Model by decoupling all scalar states
    except one Higgs doublet and the complex gauge singlet. The mass
    spectrum of the resulting singlet-extended Standard Model is
    calculated from two different perspectives: 1) using a 
    matching of the scalar sectors at next-to-leading order; 
    2) using the simplified-model approach of calculating the masses in the effective theory 
    at fixed order at the weak scale
    ignoring any connection to the full theory. Significant deviations
    between the two methods are found not only in the scalar sector, but also  
    properties of the additional fermions can be very different. Thus, only a small part of the 
    parameter space of the simplified model can be embedded in a well motivated SUSY framework. 
\end{abstract}

\maketitle

\section{Introduction}
\label{sec:intro}
The discovery of the Higgs boson in 2012 with a mass of about $m_h\approx \unit[125]{GeV}$ 
\cite{Chatrchyan:2012ufa,Aad:2012tfa,Aad:2015zhl}
strengthens the question whether additional light spin-0 fields may exist in nature.
Scalar fields transforming as singlets under the Standard Model (SM) gauge
group appear in a variety of models beyond the SM (BSM). They can serve as mediators to
hidden sectors or as dark matter (DM) candidates. In
addition, their vacuum expectation value $v$ (VEV) can help to
understand the appearance of dimensionful parameters. However, 
models predicting singlet fields can originate
from different theoretical motivations, formulated within different
frameworks such as gauge-, \mbox{gravity-,} or anomaly-mediated supersymmetry (SUSY)
breaking, extra dimensions, conformal field
theory, compositeness etc. as well as admixtures. 
Thus, the hypothetical discovery of a
singlet scalar would not necessarily point to a specific class of models.
On the other hand, there are increasing experimental constraints on e.g.
colour-charged BSM fields \cite{Sirunyan:2018vjp} that may appear along the weaker constrained 
singlet states. 
\par
Therefore, a common approach to study singlet extensions is to be as generic as possible:
after integrating out all heavy degrees of freedom, the resulting low-energy effective field theory (EFT)
is cosidered without any connection to the fundamental theory. In this way,
parameter scans of the 'simplified model' can cover many different classes of theories. 
However, this ansatz neglects not only possible correlations among the parameters, but also 
might include parameter regions that are not accessible by any reasonable full
theory. Moreover, it is not 
even clear if all other BSM particles predicted by the ultra-violet (UV) theory 
but the singlet can be decoupled in a consistent way. Therefore, it is
interesting to ask, which part of the parameter space of a simplified model is
accessible when assuming a concrete UV completion. In order to address this
question, a precise matching of the two theories as well as the evaluation 
of the renormalisation group equations (RGEs) is required. 
\par
In this context, we are going to consider the singlet extended SM (SSM) as EFT
of the Next-to-Minimal Supersymmetric Standard Model (NMSSM).
This constellation is a new variant of the popular ideas of high-scale or Split SUSY
\cite{Wells:2003tf} which usually consider the Minimal Supersymmetric Standard Model (MSSM)
with a specific R-symmetry-breaking pattern \cite{ArkaniHamed:2004yi,ArkaniHamed:2004fb} as full theory in the UV. 
Split SUSY has the advantage of providing a dark matter candidate and improving the unification 
of the SM gauge couplings compared to low-scale SUSY scenarios
\cite{Giudice:2004tc}. In addition, non-minimal Split SUSY can be
connected to cosmological observables such as baryon asymmetry
\cite{Demidov:2006zz,Demidov:2016wcv} or gravitational waves
\cite{Demidov:2017lzf}.
As we will discuss, the considered singlet extension of the MSSM does not allow
to decouple the electroweakinos while keeping the singlet light,
which is why a purely scalar-singlet extension of the SM can only
hardly be motivated by non-minimal SUSY.
\par
In general, the presence of additional heavy states in the UV theory makes the inclusion of higher-order
corrections to the matching conditions mandatory in order to keep the theoretical uncertainties under control \cite{Vega:2015fna,Athron:2016fuq,Bagnaschi:2017xid,Staub:2017jnp,Allanach:2018fif}. 
The higher-order effects in the matching of the high-scale MSSM, where all BSM
states are very heavy, to the SM can alter the Higgs mass by several GeV \cite{Allanach:2018fif}. 
Moreover, one needs to include carefully the effects of potentially light states. This has been discussed for instance 
in the context of the Two Higgs Doublet Model as a low-energy
theory of the MSSM \cite{Haber:1993an,Beneke:2008wj,Gorbahn:2009pp,Lee:2015uza, Bahl:2018jom,Gabelmann:2018axh,Bahl:2019ccs}: it was shown that the SM-like
Higgs boson mass prediction using a proper
decoupling of heavy -- and only heavy -- scalars can differ by up to \unit[10]{GeV} compared to
simple high-scale SUSY approaches which also treat the second Higgs doublet as
if it would be heavy. 
On the other side, the impact of higher-dimensional operators is usually
sub-dominant for the Higgs boson mass prediction. 
It was shown for the MSSM that the impact of dimension-six
operators, scaling with $\order{\nicefrac{\vev^2}{\msusy^2}}$, on the Higgs
boson mass becomes very weak if the BSM scale is above \unit[1-2]{TeV} \cite{Bagnaschi:2017xid}. 
Thus contributions of higher-dimensional
operators are negligible for this work, because we require not only the SM VEV to be smaller than the
decoupling scale, $\vev\ll \msusy $, but also the VEV of the additional (light) 
 singlet $\vs$ must be comparable to the singlet mass to get a consistent
 effective theory and a stable potential.
\par
The remainder of the paper is organised as follows. In \cref{sec:emergence}, the NMSSM is
introduced focusing on the soft SUSY breaking scalar part as well as the
resulting properties of the low-energy model, the SSM. \cref{sec:matching} discusses higher-order effects 
that arise when couplings with positive mass dimension are involved in the matching. 
In \cref{sec:results} we compare the simplified-model with 
the matching approach. Conclusions and
outlooks are given in \cref{sec:conclusion}.

\section{How does a SSM emerge from SUSY?}
\label{sec:emergence}
Simplified models that involve additional scalars are often motivated by SUSY because 
a consistent and phenomenological viable SUSY model needs at least one additional doublet. 
Thus, simplified models assume that  SUSY contributes with only very 
few new degrees of freedom at the weak scale, but all other positive 
aspects of the full theory, like diminishing the hierarchy problem, come into play at 
a higher scale \cite{Alves:2011wf,Contino:2013kra,Costa:2014qga,Buttazzo:2015bka,Costa:2015llh,Borschensky:2018zmq}.
However, this means that all additional states predicted by SUSY need to decouple in a
consistent way. As we will show, this is not always possible since 
assumptions such as the introduction of discrete symmetries in the scalar
sector are not compatible with those predicted by SUSY and specific soft-SUSY-breaking patterns. In
addition, the inclusion of higher-order corrections shows that it is not
possible to enforce a large mass hierarchy 
between specific states without decoupling the singlet as well. We want to discuss this 
at the example of matching the NMSSM to the SSM.
\subsection{Possible low-energy limits from the NMSSM}
There is a rich collection of motivations and introductions for the softly broken (N)MSSM
available in the literature, see for instance Refs.~\cite{Martin:1997ns,Maniatis:2009re,Ellwanger:2009dp} and references therein. 
Therefore, we skip the motivation and continue with
the discussion of how to obtain a low-energy limit of the NMSSM 
that involves (at least) one scalar gauge singlet. 
The most general\footnote{The absence of a tadpole term in \cref{eq:emergence:wuv} is due to a redefinition of the singlet field such that only a soft-breaking tadpole in \cref{eq:LSB_nmssm} remains.} superpotential of the NMSSM without assuming any global symmetry is
\begin{equation}
\label{eq:emergence:wuv}
\Wsuper_\text{NMSSM} =  {\Wsuper_\text{MSSM}}  + M_S \hat{S}^2 + \lambda\hat{S}\hat{H_u}\ccdot\hat{ H_d} + \frac{\kappa}{3}\hat{S}^3 ,
\end{equation}
with 
\begin{equation}
\label{eq:wmssm}
    \Wsuper_\text{MSSM} =  \mu \hat{H_u}\ccdot\hat{ H_d} + Y_u \hat{H}_u \ccdot \hat{Q} \hat {u} + Y_d \hat{H}_d \ccdot \hat{Q} \hat {d} + Y_e \hat{H}_e \ccdot \hat{L} \hat {e},
\end{equation}
where we follow the notation of \cite{Goodsell:2014pla}.
The corresponding soft-SUSY breaking terms are
\begin{equation}
\label{eq:LSB_nmssm}
    L_\text{NMSSM}^\text{soft} =  L_\text{MSSM}^\text{soft} + t {S} + B_S {S}^2 + T_\lambda {S} {H_u}\ccdot { H_d} + \frac{T_\kappa}{3} {S}^3 ,\\
\end{equation}
with 
\begin{equation}
\label{eq:LSB_mssm}
    L_\text{MSSM}^\text{soft} =  B_\mu {H_u}\ccdot{H_d} + T_u {H}_u \ccdot \tilde{Q} \tilde {u} + T_d {H}_d \ccdot \tilde{Q} \tilde{d} + T_e {H}_e \ccdot \tilde{L} \tilde{e}
\end{equation}
and the soft-breaking scalar mass squared terms $m_\phi^2 |\phi|^2$ for all scalars
$\phi = \{H_d, H_u, S, \tilde{Q},\tilde{u},\tilde{d},\tilde{L},\tilde{e}\}$ as
well as soft fermion masses $M_{\lambda=1,2,3}$ for the bino ($\tilde{B}$), wino
($\tilde{W}$) and the gluino ($\tilde{g}$). 
\par
\begin{figure}[t]
    \includegraphics[width=0.4\linewidth]{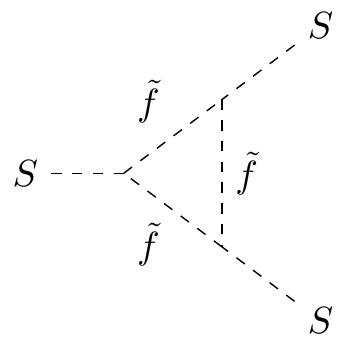}
    \includegraphics[width=0.4\linewidth]{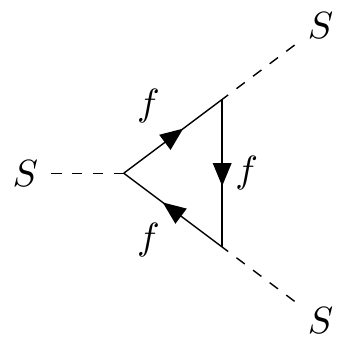}
    \caption{Generic one-loop contribution of a scalar (fermion) $\tilde{f}$
    ($f$) to the trilinear scalar self coupling. While the scalar contribution
scales as $\nicefrac{\left[\text{Trilinear coupling}\right]^3}{m_{\tilde{f}}^2}
\xrightarrow{m_{\tilde{f}}\to \infty} 0$, the fermionic
contribution,$\left[\text{Yukawa coupling}\right]^3 m_f$, does not decouple in
the limit $m_f\to \infty$.}
    \label{fig:triangleoneloop}
\end{figure}

The low-energy theory shall contain a complex singlet. The most general scalar
potential involving a complex singlet is 
\begin{align}
\label{eq:split:left:vgauge}
    V(H,S) = &\,\, m_H^2 HH^\dagger + \frac{\lambda_H}{2} \left|HH^\dagger\right|^2 \nonumber \\
           + &\,\, m_S^2 |S|^2 + (\tilde{m}_S^2 S^2 + \text{h.c.}) + \nonumber \\
           + &\,\, \frac{\lambda_S}{2} |S|^4 + (\lambda'_S |S|^2 S^2 + \lambda''_S S^4 + \text{h.c.})   \\
           + &\,\, \lambda_{SH} |S|^2 HH^\dagger + (\lambda'_{SH} S^2 H H^\dagger + \text{h.c.}) \nonumber \\
           + &\,\, (\kappa_{SH} SHH^\dagger + \kappa_S SSS + \kappa'_S |S|^2 S  + \,\,\text{h.c.}\,). \nonumber
\end{align}
For consistency reasons (such as vacuum stability) all dimensionful parameters in this potential as well
as the two VEVs $v$ and $v_S$ of the doublet $H$ and the singlet $S$ must be roughly of the same size. It seems that 
\cref{eq:emergence:wuv,eq:wmssm,eq:LSB_nmssm,eq:LSB_mssm} provide enough freedom to decouple all squarks, sleptons,  the second doublet,  electroweakinos, and the singlino. However, this 
is only correct at leading order. At one-loop order, trilinear self-couplings
would receive corrections from heavy fermions and scalars depicted in \cref{fig:triangleoneloop}.
For instance, the trilinear self-coupling $\kappa_S$ 
will receive loop corrections at the matching scale involving the singlino
($f=\tilde{S}$ in the right diagram of \cref{fig:triangleoneloop}) that scale with
\begin{equation}
    \label{eq:kappanondecouple}
    \kappa_S^\text{one loop} \sim \kappa^3\,  m_{\tilde{S}}, 
\end{equation}
where $\kappa$ and $m_{\tilde{S}}$ is the Yukawa
coupling and mass of the Singlino. Therefore, $\tilde{S}$ would not decouple
from the low-energy Lagrangian in the limit $m_{\tilde{S}}\to \infty$.
Similarly, the coupling $\kappa_{SH}$ receives loop corrections from
Higgsino/gaugino loops, e.g. 
\begin{align}
    \kappa_{SH}^\text{one loop} \sim &\,\, g_i^2\, \lambda\, \left( M_i +
        m_{\tilde{H}_{u,d}} \right)\label{eq:kappanondecouple2} \\ 
                                     &+ \kappa\,\lambda^2\, \left( m_{\tilde{S}} + m_{\tilde{H}_{u,d}} \right)\, ,
    \,\,i=\{1,2\} ,\nonumber
\end{align}
where $\lambda$ is the singlet-Higgsino-Higgsino coupling, and $g_i$ the
Higgs-Higgsino-gaugino coupling.

Thus, all additional SUSY fermions that couple to the singlet must have masses similar to the scalar
singlet in order to ensure trilinear couplings of the order of the low-energy
scale.
On the other hand, scalar contributions to trilinear couplings, shown in the
left diagram of \cref{fig:triangleoneloop}, always decouple. Therefore, the second doublet 
as well as the squarks and sleptons can be kept heavy by assuming 
the soft-squared parameters $B_\mu$ and $m_{\tilde f}^2$ ,$\tilde{f} =
\{\tilde{Q},\tilde{U},\tilde{e},\tilde{d},\tilde{u}\}$ to be much larger than
the electroweak scale. 
\par
Neglecting for the moment the effect of singlet-doublet mixing
we can easily calculate the masses $m_{\phi_s}^2$, $m_{\sigma}^2$ of the
CP-even and -odd component of the singlet. In the limit $t,v\to 0$ these are given by 
\begin{align}
m_{\phi_s}^2 =& \frac{3}{\sqrt{2}} \kappa  M_S \vs + \frac{1}{\sqrt{2}} \vs
T_\kappa + 2 \kappa^2 \vs^2\, \\
m_{\sigma}^2 =& -2 B_S - \frac{1}{\sqrt{2}} v_S (\kappa  M_S + 3
                T_\kappa) \;.
\end{align}
Thus, taking $|B_S|$ to be very large, too, would also decouple
the CP-odd component of the singlet and result in the SM extended by a real singlet
as EFT. However, we are going to concentrate on the complex-singlet case.
Based on these considerations, the hierarchy of dimensionful parameters used in the following can be
summarised using two different scales:
\begin{align}
    \label{eq:emergence:tscaling}
    \sqrt{B_S},t,\mu,\vs,T_{\lambda,\kappa,\tilde{e},\tilde{d},\tilde{u}}, M_{S,1,2,3} \sim  m_\text{\tiny Sing} \sim  \unit[1]{TeV}, \\
    \label{eq:emergence:m2scaling}
    B_\mu,\,m_{\tilde{l},\tilde{q},\tilde{u},\tilde{e},\tilde{d}}^2 \sim  \,  \msusy^2 \gg \left(\unit[1]{TeV}\right)^2\, .
\end{align}
In order to reduce the number of free parameters, we assume $t=M_S=\mu = B_S = 0$ for
simplicity, which leads to a scale-invariant superpotential without a
tadpole\footnote{A non-zero tadpole $t$ would also lead to a singlet mass
that scales with $\nicefrac{t}{\vs}$ which would decouple the singlet for small
values of $\vs$.} problem \cite{Ellwanger:2009dp}.
There is -to our knowledge- no discrete (R-)symmetry that would
justify setting all mentioned parameters to zero simultaneously.
However, these kind of parameters would anyways be neglected in the
matching of our calculation, since they are suppressed by factors of
$\msusy^{-2}$.
Furthermore, $\mu$ is the only parameter that would lead to a new
operator $\mu \tilde{H}_u \tilde{H}_d$  in the EFT in eq. (20), which
we do not consider because it mainly concerns the Higgsino masses but
has minor effect on the singlino and scalar sector onto which we
focus.
For further simplification, we assume also the soft-sfermion mass matrices and trilinear
couplings to be degenerate and flavour-diagonal
\begin{equation}
    \label{eq:emergence:sqdegenerate}
    \begin{aligned}
        T_{\tilde{i}}     & = Y_i A_0,\,\,        A_0\approx\order{m_\text{Sing}},\,\, i={u,d,e},\\
        m_i^2   & =\mathbb{1} \, \msusy^2,\,\, i={\tilde{l},\tilde{q},\tilde{u},\tilde{e},\tilde{d}}
    \end{aligned}
\end{equation}
and only take the third-generation Yukawa couplings $Y_i^{(3,3)}$ into account. 
The additional Higgs boson states are automatically degenerate in the mass $m_A$ in the decoupling limit:
\begin{equation}
    \label{eq:emergence:bmu}
    m_A^2 \xrightarrow{B_\mu\gg m_W^2} \frac{1+\tan^2\beta}{\tan\beta} B_\mu\, .
\end{equation}
With \cref{eq:emergence:m2scaling} it follows that 
$m_A^2 \sim \order{\msusy^2}$, i.e. all heavy scalars have masses of the same
size and can be decoupled at one matching scale in a consistent way.
Thus, there are two additional BSM scales: $\msusy$ in the UV and one
intermediate mass scale $m_\text{\tiny Sing}\ll\msusy$ in
the EFT. The latter can either be smaller or larger than
$\vev$. However, $m_\text{\tiny Sing}$
should not be much larger than \unit[1]{TeV} in order to keep it close to the
SM scale and ensure the validity of the EFT approach.

\subsection{The low-energy Lagrangian}
As a result of
\cref{eq:emergence:tscaling,eq:emergence:m2scaling,eq:emergence:sqdegenerate,eq:emergence:bmu} 
only fermions, one Higgs doublet and one complex singlet do not decouple.
Thus, the considered EFT
consists of the SM complemented with a complex singlet scalar $S$ as well as
Weyl fermions transforming as SU(2) singlets $\tS$ (singlino),
$\tB$ (Bino), a pair of two
SU(2)xU(1) doublets  $\tHu$, $\tHd$ (Higgsinos), a SU(2) triplet
$\tW$ (Wino) and an SU(3) octet $\tG$ (gluino).

To avoid confusion, we make use of the same notation for the fields in the EFT
as we did in the NMSSM although they differ by a field renormalization. In order 
to simplify the following discussion, it turns out to be helpful to 
rewrite the complex singlet into a pair of real CP-even and -odd components
\begin{equation}
    S=\frac{1}{\sqrt{2}}\left(S_r + i S_i\right) \, , \,\,\, \langle S_r \rangle=\vs.
\end{equation}
One observes that they couple differently to the Higgs doublet as well as the
fermion sector. For instance, the coupling $\kappa_{SH}$ only couples the
CP-even part to the doublet, but not the CP-odd part (which would be
a CP-violating interaction). Thus, the CP-even/odd interactions will behave
differently under the RGE running and result in different couplings for the
CP-even/odd components in the low-energy Lagrangian. 
In this basis \cref{eq:split:left:vgauge} becomes
\begin{equation}
    \label{eq:split:left:split}
    \begin{aligned}
        V(H,S) \equiv  & \,\,V(H,\Si,\Sr) \\
                   =  & \,\, m_H^2 HH^\dagger + \frac{\lambda_H}{2}
                   \left(HH^\dagger\right)^2 + \frac{m_{\Sr}^2}{2} \Sr\Sr +\\
                   + &\,\, \frac{\lambda_{\Sr}}{8} \left(\Sr\Sr\right)^2
        +  \frac{m_{\Si}^2}{2} \Si\Si + \frac{\lambda_{\Si}}{8} \left(\Si\Si\right)^2 \\
        + &\,\, \frac{\lambda_{S_{ri}}}{4} \Sr\Sr\Si\Si
                    +  \frac{\lambda_{{SH}_r}}{2} \Sr \Sr H H^\dagger  \\
                    + &\,\, \frac{\lambda_{{SH}_i}}{2} \Si \Si HH^\dagger + \sqrt{2} \kappa_{{SH}_r} \Sr H H^\dagger \\
                      -&\,\, \frac{3}{\sqrt{2}} \kappa_{S_{ri}} \Sr\Si\Si +
                      \frac{1}{\sqrt{2}} \kappa_{\Sr} \Sr\Sr\Sr \;.
    \end{aligned}
\end{equation}
Using this parametrisation we do not only account correctly for RGE effects,
but also include operators like $S |S|^2+h.c$\footnote{This operator is actually linearly dependent to $\kappa_{S_r}$ and $\kappa_{S_{ri}}$. Likewise, all other \textit{primed} couplings in \cref{eq:split:left:vgauge} are linearly dependent.}, that are non-existent in the
NMSSM at tree-level but generated at the one-loop
order. This is because a matching of \cref{eq:split:left:split} already involves the most
general CP-conserving potential. \\
The 2x2 mass matrix of the CP-even eigenstates reads
\begin{equation}
    \label{eq:mhh2}
    \textbf{m}_H^2 =
    \begin{pmatrix}
        m^2_{11} & m^2_{12} \\
        m^2_{12} & m^2_{22}
    \end{pmatrix},
\end{equation}
with the components
\begin{equation}
    \label{eq:mhh}
    \begin{aligned}
        m^2_{11} &= \lambda_H\, \vev^2 \, , \\
        m^2_{12} &= \sqrt{2}\, \vev\, \kappa_{{SH}_r} + \vev\, \vs\,\lambda_{{SH}_r}\,\,\, \text{and} \\
        m^2_{22} &= -\frac{\vev^2\, \kappa_{{SH}_r}}{\sqrt{2}\, \vs} + \frac{3\, \vs\,
        \kappa_{S_r}}{\sqrt{2}} + \vs^2\, \lambda_{S_r} \, ,
    \end{aligned}
\end{equation}
where tree-level tadpole conditions have been used to eliminate $m_H^2$ and
$m_{S_r}^2$.
The eigenvalues $m_h^2$/$m_s^2$ of $\textbf{m}_H^2$ are associated with the squared 
masses of the doublet-/single-like mass eigenstates. 
The CP-odd state has the mass
\begin{equation}
    \label{eq:split:mawithoutmatch}
    m_a^2 = m_{\Si}^2 + \frac{\lambda_{\Si}^2 \vev^2}{2}-3\sqrt{2}\, \kappa_{S_{ri}}\vs + \frac{\lambda_{S_{ri}} \vs^2}{2}.
\end{equation}
Since we did not assume a CP-violating vacuum, the $m_{\Si}^2$ contribution in
\cref{eq:split:mawithoutmatch} cannot be eliminated (in contrast to $m_{\Sr}^2$). However,
if we neglect for a moment the RGE running to the matching scale, we can
identify $m_{\Si}^2$ with $m_{\Sr}^2$ before using the tadpole
equations\footnote{This will be discussed in more detail later, see \cref{eq:bcslow}} and find 
\begin{equation}
    \label{eq:split:ma}
    m_{a}^2 \xrightarrow{\msusy\to \vev} -\frac{ 9\,\vs^2\, \kappa_S + \vev^2\, \kappa_{SH}}{\sqrt{2}\,\vs}\, ,
\end{equation}
which shows that complex singlet extensions with a $\mathbb{Z}_2$ symmetry,
i.e. $\kappa_S,\kappa_{SH}\to 0$, suffer from a massless pseudoscalar goldstone boson. 
\par
In the fermionic sector we find the 
following mass as well as interaction terms with scalars: 
\begin{equation}
    \label{eq:split:lino}
    \begin{aligned}
        \Lag_\text{fermion} = & \,\,  Y_d^{\overline{\text{MS}}}\, q\, H^\dagger\, d -
                                Y_u^{\overline{\text{MS}}} \,q\, H  - Y_e^{\overline{\text{MS}}}\, l \,H\, e \\
                            - & \,\,  g_2^u \tHu\, H^\dagger\, \tW  -
                                \frac{g_1^u}{\sqrt{2}}\, \tHu\, H^\dagger\, \tB   \\
                            - & \,\,  g_2^d\, \tHd\, H\, \tW  -
                                \frac{g_1^d}{\sqrt{2}}\, \tHd\, H\, \tB   \\
                            - & Y_S^u\, \tS\, \tHu\, H^\dagger  - Y_S^d\, \tS \, \tHd\, H \\
                            - & \,\,  Y_{\tS}\, {\tS}\, {\tS}\, {S} -
                                \frac{Y_{ud}}{\sqrt{2}}\, S\, \tHu\ccdot \tHd \\
                            - & \,\,  \frac{M_{\tG }}{2}{\tG}{\tG} - \frac{M_{\tB}}{2} \tB\tB - \frac{M_{\tW}}{2} \tW\tW \\
                            + & \,\,  \text{h.c.} \, ,
    \end{aligned}
\end{equation}
which is analogous to the Split MSSM Lagrangian in Ref. \cite{Bagnaschi:2014rsa} extended by
a singlet fermion $\tilde{S}$ and scalar S. The splitting of $S$ into its CP-even/odd part and the introduction of
independent couplings was also performed in the Yukawa sector. However, the
different Yukawa couplings are not of particular interest for this letter. To keep the
notation simple, we always refer to the Yukawa coupling to the CP-even
singlet component, if not stated elsewhere.
It is important to stress, that all Lagrangian parameters are 
independent of each other from the view-point of the simplified model, 
even if the notation indicates their relation
given above the matching scale. In order to sum up this part, we show in 
\cref{tab:parametercounting} the five free parameters
(excluding the VEVs) contributing at tree-level to the light scalar masses in
the NMSSM as well as in the SSM.

\begin{table}[tb]
        \begin{tabular}{l|cc}
            &                                                                           & $\Sigma$ \\ \hline
        UV  &  $\lambda,\, \kappa,\, T_\lambda,\, T_\kappa,\, \tan\beta$                & 5 \\
        EFT & $\lambda_{H},\, \lambda_{S},\, \lambda_{SH},\, \kappa_{S},\, \kappa_{SH}$ & 5 \\
        \end{tabular}
        \caption{Parameter counting contributing to the light scalar masses in the
        (Split) NMSSM as well as the SSM.}
        \label{tab:parametercounting}
\end{table}

\subsection{One-Loop Matching}
At the matching scale $\msusy$, the fields $S_r$ and $S_i$ are contained in the
complex singlet $S$. Thus, they are degenerate in mass and couple identically. 
Therefore, we have the following additional \textit{matching} relations
\begin{equation}
    \label{eq:bcslow}
    \begin{aligned}
        \kappa_{{SH}} \equiv &\,\, \kappa_{{SH}_r} \,\, , \\
        \kappa_{S}      \equiv &\,\, \kappa_{\Sr} =  \kappa_{S_{ri}}\,\, ,\\
        \lambda_{S}     \equiv &\,\, \lambda_{S_{ri}} = \lambda_{\Si} = \lambda_{\Sr}\,\, ,  \\
        \lambda_{SH}  \equiv &\,\, \lambda_{SH_{r}} = \lambda_{SH_i}\,\,
        \text{and}  \\
         m_{S}^2        \equiv &\,\, m_{\Sr}^2 = m_{\Si}^2\, ,
    \end{aligned}
\end{equation}
which restore the complex-singlet structure at the matching scale.
Thus, the information about all heavy SUSY states is encoded in a small set
of effective parameters $\kappa_{{SH}}$, $\kappa_{S}$, $\lambda_{S}$,
$\lambda_{SH} $ and $\lambda_H$ which we need to calculate as
precisely as possible. 
It was already mentioned in 
the introduction that a tree-level matching is no longer sufficient because of
the precise LHC measurements of the Higgs boson mass. For this reason, we use the
computer tool \SARAH (version 4.14.1)
\cite{Staub:2009bi,Staub:2010jh,Staub:2012pb,Staub:2013tta}, which was
extended to be able to perform a matching of the scalar
sector of two theories at next-to-leading order, see
Refs.~\cite{Gabelmann:2018axh,Braathen:2018htl} for more details. 
\par
The scalar quartic and trilinear effective couplings are given through NMSSM
$D$- and $F$-terms as well as one-loop corrections at the matching scale $\msusy$:
\begin{align}
    \label{eq:split:vmatch}
    \lambda_H =& \frac{1}{4} (g_2^2 + g_1^2) \cos^22\beta +
    \frac{1}{2}\lambda^2\sin^22\beta + \delta^{(1)} \lambda_H\, , \nonumber \\
    \lambda_S =& 2 \kappa^2 + \delta^{(1)} \lambda_S \, ,\nonumber \\
    \lambda_{{S}{H}} =& \lambda^2 - \frac{2\,\kappa\,\lambda\,\tb}{1+\tb^2}
    - \frac{T_\lambda^2}{m_A^2}\left(\frac{\tb^2-1}{\tb^2+1}\right)^2 + \delta^{(1)} \lambda_{SH}\, ,\nonumber \\
    \kappa_S  =& \frac{1}{3} T_\kappa + \delta^{(1)} \kappa_S\, ,\nonumber \\
    \kappa_{SH}  =& - T_\lambda \frac{\tb}{1+\tb^2}+ \delta^{(1)} \kappa_{SH}\, ,
\end{align}
where the short-hand notation $\tb=\tan\beta$ is used. The coupling $\lambda_{SH}$ receives additional
non-local contributions from tree-level diagrams with one internal heavy
Higgs boson shown in the left diagram of \cref{fig:oneloopstops}.  
The $\delta^{(1)}$-terms denote the one-loop corrections to the shown
tree-level contributions. However, the expressions are rather lengthy while 
the calculation using \SARAH is relatively simple and can be reproduced on 
any standard personal computer. Thus, we refrain from showing them
here\footnote{The expressions can easily be obtained with \SARAH using
  the file \textsl{Models/SMSSM/Matching\_SplitSUSY.m} included in the
  public package.}. 
\par
The one-loop matching of Yukawa couplings is only required for the Higgs boson mass calculation
 when a running above the NMSSM scale $\msusy$ is
 considered. However, 
we will not consider such a scenario here, i.e.  only the tree-level matching
conditions are required:
\begin{equation}
\label{eq:split:yukawamatch}
\begin{aligned}
   g_1^u &=  g_1 \sin\beta\,, \,\,& g_2^u &= g_2 \sin\beta \,,\\
   g_1^d &=  g_1 \cos\beta\,, \,\,& g_2^d &= g_2 \cos\beta \,,\\
   Y_{ud} &= \lambda                        \,, \,\,& Y_{\tS} &= \kappa \,,\\
   Y_{S}^d &= \lambda \sin\beta             \,, \,\,& Y_{S}^u &= \lambda \cos\beta \,,\\
   Y_{e,d}^{\overline{\text{MS}}} &= \cos\beta\, Y_{e,d}^{\overline{\text{DR}}}
   \,   ,\,\,& Y_u^{\overline{\text{MS}}} &=
   \sin\beta\, Y_u^{\overline{\text{DR}}} \,,\\
   M_{{\tB},{\tW},{\tG}} &= M_{1,2,3}             \,       .     &         &
\end{aligned}
\end{equation}
\par
\begin{figure}[h]
    \centering
    \includegraphics[width=0.85\linewidth]{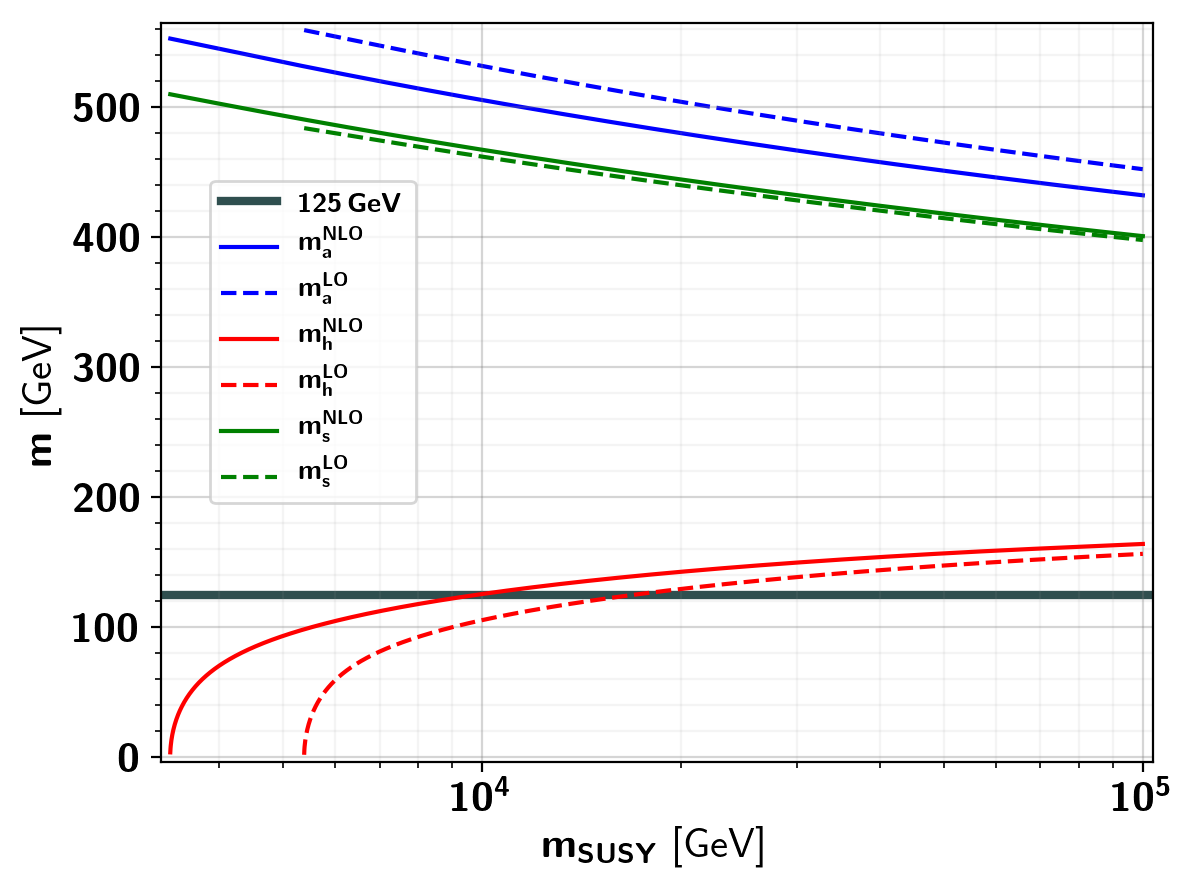}
    \caption{Example for the scalar spectrum of the Split NMSSM. The red
    straight (dashed) line shows the doublet-like Higgs boson mass as a
function of the matching scale using NLO (LO) matching conditions. The mass
associated with the CP-even (odd) state is shown in green (blue). The horizontal
line corresponds to the central value of the measured Higgs boson mass. The
used parameter values are: $M_1:M_2:M_3=0.15:1:\unit[3]{TeV},\,
\tan\beta=5,\, \lambda=\nicefrac{\kappa}{2}=0.9,\,
T_\lambda=-T_\kappa=\unit[500]{GeV},\, A_0=\vs=\unit[100]{GeV}\,
\text{and}\, m_A=0.1\msusy$.}
\label{fig:massesexample}
\end{figure}
We give a brief example for the mass spectrum and the NLO effects in \cref{fig:massesexample} 
where we plot the scalar mass spectrum calculated in the
EFT at the scale $\vev$ as function of $\msusy$. In a first step of the
calculation  a running of the effective couplings to the scale $\msusy$ is carried out,
before the matching is performed using \cref{eq:split:yukawamatch,eq:split:vmatch}. Finally, 
the RGEs run back to $\vev$ and the effective Higgs boson masses are computed at the low-scale
at two-loop order\footnote{See Fig. 5 of Ref. \cite{Gabelmann:2018axh} for a
detailed discussion on the effective Higgs mass calculation.}. All numerical calculations are performed with a \SARAH generated \SPheno \cite{Porod:2003um,Porod:2011nf}  version
which includes the full two-loop RGEs. 
One can clearly see the significant impact of the NLO corrections: at LO the desired Higgs mass 
of about \unit[125]{GeV} is obtained for $\msusy$ larger than \unit[10]{TeV}
, while NLO corrections can
pull $\msusy$ below \unit[10]{TeV}.
The impact of higher-order corrections from the matching conditions on the mass
spectrum is discussed in further detail in the next section. 

\section{Matching of Mass Parameters}
\label{sec:matching}
The previous section showed that higher-order corrections to matching
conditions can have a significant impact on the soft-SUSY-breaking
scale required for being consistent with the Higgs boson mass measurement. In this
section we further examine the behaviour of these corrections.
\par
\begin{figure}[tb]
    \centering
    $\delta^{(1)} \kappa_{S} \supset \,$
    \parbox[c]{2.5cm}{
    \includegraphics[width=0.2\textwidth]{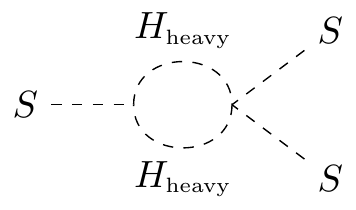}\\
     }
    $\qquad\,\,\, \xrightarrow[\to\infty]{m_\text{\tiny heavy}} -\frac{\kappa\,\lambda\, T_\lambda \cos^2 2\beta}{8 \pi^2}$
    \caption{One-loop contribution to the matching condition of the trilinear
        singlet self coupling which does not vanish if $\msusy\to\infty$.
    In this example all masses of the heavy doublet $H_\text{\tiny heavy}$ are assumed to be degenerate at the matching scale $\msusy$. 
All other contributions (i.e. wavefunction renormalisation and triangle diagrams) decouple in this limit.}
    \label{fig:tridiag}
\end{figure}

Quartic scalar couplings are dimensionless in $D=4$ dimensions. Thus, the
dependence on the mass scale involved in the matching, $\msusy$,
is guaranteed to disappear in the limit $\msusy\to\infty$.
However, this is not the case for parameters with mass dimension greater
than zero, which appear in the Lagrangian of the SSM, cf.
\cref{eq:split:left:split} and the discussion of \cref{eq:kappanondecouple,eq:kappanondecouple2}.
A related behaviour can be seen explicitly by looking at the one-loop 
correction to the trilinear
self-coupling $\kappa_S$ at the matching scale,  
depicted in \cref{fig:tridiag}. 
The contribution from a
heavy Higgs boson in the loop does not decouple, although being independent of
the Higgs mass itself, but becomes 
\begin{equation}
    \label{eq:kappa1l}
\delta^{(1)} \kappa_{S} \simeq   -\frac{\kappa\,\lambda\, T_\lambda \cos^2 2\beta}{8 \pi^2},
\end{equation}
which can easily reach several GeV if $T_\lambda\approx
\unit[1]{TeV}$. As long as the loop corrections are smaller than the tree-level contributions,
one might not worry about this behaviour. However, there are various situations 
which can lead to surprisingly large, but not necessarily unphysical, higher-order corrections to the scalar
masses and trilinear couplings. One interesting case is $T_\kappa
\simeq 0$, which could either be 
 due to some additional symmetry or just a numerical coincidence. In both cases one would find
$\kappa_{S} \simeq 0$ at tree-level at the matching scale. However, if it is just a coincidence or if the symmetry is broken by loop corrections, 
one would find $|\delta^{(1)} \kappa_{S}|\gg 0$ if $T_\lambda \neq 0$. Similarly, an accidental symmetry or numerical coincidence
could cause a cancellation within the pseudo-scalar mass given in \cref{eq:split:mawithoutmatch,eq:split:ma}, i.e. $a$ would 
be nearly massless at tree-level but could receive a sizeable mass from different higher-order corrections to
$\kappa_{S}$ and $\kappa_{SH}$. Another factor in this discussion is the
RGE running that spoils the identities in \cref{eq:bcslow} after the matching
has been applied. This effect can also lead to an accidental enhancement of higher-order
matching conditions if, for instance, a one-loop matching changes the sign of a
coupling w.r.t the tree-level matching.
\par
In principle, one could avoid the entire discussion about
large higher-order corrections by using a renormalisation scheme different from
the minimal subtraction ($\overline{\text{MS}}$) scheme.
In the case of an on-shell (OS) scheme, the masses are fixed to all orders by adding fine-tuned counter-terms \cite{Braathen:2019pxr}. 
However, one might expect that large loop corrections then show up at another place. For instance,
large OS counter terms will influence the two-loop corrections in the doublet
sector which itself has only very restricted freedom concerning the mass counter
terms. Even more important, one would not be allowed to feed these
parameters into the RGEs, which are defined for 
$\overline{\text{MS}}$ parameters only. Therefore, the proper matching is --by
now-- the most suitable 
solution to relate the SSM and NMSSM in the most predictive way.
\par
In what follows, we continue with a quantitative discussion of the higher-order
corrections. In this context, we do not assume any 
additional symmetries which would predict tiny or zero $T_\kappa$ or $m_a$, 
respectively. However, we would like to check how likely parameter configurations at the matching scale appear 
where the NLO correction supersede the LO contributions. 
\begin{table}[tb]
    \centering
    \begin{tabular}[t]{ll}
        parameter              & scan range                     \\ \hline
        $\lambda$              & [-3, 3]                    \\
        $\kappa$               & [-3, 3]                    \\
        $\tan\beta$            & [1, 50]                    \\
        $\msusy$ & [$10^3$, $10^{16}$] GeV   \\
        $m_A$ & $r\cdot \msusy$ \\
        $r$   & [0.01, 100] 
    \end{tabular}
    \begin{tabular}[t]{ll}
        parameter              & scan range                     \\ \hline
        $T_\lambda$            & [-5000, 5000] GeV                \\
        $T_\kappa$             & [-5000, 5000] GeV                 \\
        $M_{1,2}$              & [10, 3000] GeV            \\
        $M_3$                  & [1000, 3000] GeV             \\
        $\vs$                  & [0, 3000] GeV            \\
        $A_0$                  & [-500, 500] GeV
    \end{tabular}
    \caption{Scan ranges for a random scan over the NMSSM parameter space. All
    parameters are input at the matching scale $\msusy$.}
    \label{tab:split:scanrange}
\end{table}
In order to cover large parts of the parameter space, we perform a random scan
according to the ranges in \cref{tab:split:scanrange} using the generated
\SPheno code and require $m_h=\unit[(125\pm 2)]{GeV}$ for the results obtained with NLO matching
conditions. Furthermore, we require $\msusy$ and $m_A$ to be at least twice as
large as the largest mass parameter of the low-energy Lagrangian.
\par
\begin{figure}[t]
    \centering
    \includegraphics[width=0.42\textwidth]{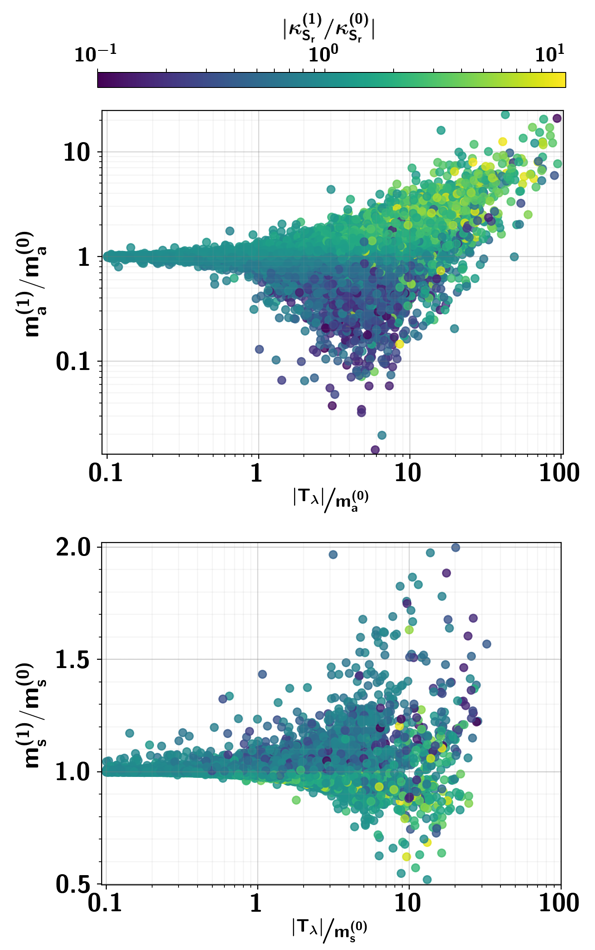}\\
    \caption{Relative size of the one-loop corrections to the mass $m_s$
        ($m_a$) of the
        (pseudo) scalar ($a$) $s$ in \% and the trilinear singlet-self coupling
        $\kappa_{\Sr}$ evaluated at the scale $\vev$ as a function of the ratio of the soft-SUSY breaking
        coupling $T_\lambda$ and the tree-level mass $m_s$ ($m_a$). It is  
        $0\lesssim |T_\lambda|,m_a^{(0)}\lesssim \unit[5]{TeV}$ and
        $0.12\lesssim m_s^{(0)}\lesssim \unit[7]{TeV}$.
    }
    \label{fig:trilinearsmams}
\end{figure}

In \cref{fig:trilinearsmams} we show the relative size of the higher-order
corrections to the scalar singlet masses $m_a$ and $m_s$ evaluated at the scale
$\vev$ when using tree-level (denoted with a $(0)$-label) or
one-loop (denoted with a $(1)$-label) matching conditions at the matching scale. 
The x-axis was chosen to show the 
$\nicefrac{T_\lambda}{m_{a,s}^{(0)}}$ dependence, because $T_\lambda$ presumably
becomes important for the one-loop shift $\delta^{(1)} \kappa_{\Sr}$ which
itself can have a large impact on the one-loop correction to the tree-level masses
of $a$ and $s$. The colour of the points indicates the relative size of
the running value of $\kappa_{S_r}$ at $\vev$ between a tree-level and one-loop matching
at $\msusy$. For $m_a$ a clear correlation
with $\nicefrac{T_\lambda}{m_a^{(0)}}$ and
$\nicefrac{\kappa_{\Sr}^{(1)}}{\kappa_{\Sr}^{(0)}}$ is visible. Even for large
values of $T_\lambda$ a relatively small $m_a$ can be realized at tree-level by
choosing $T_\kappa$ accordingly, cf. \cref{eq:split:ma,eq:split:vmatch}.
However, at the one-loop order the contributions of \cref{fig:tridiag} scale with
$T_\lambda$ and can push the loop-corrected masses to up to ten times larger
masses. 
\par
A similar behaviour is observed for $m_s$ in the lower plot of
\cref{fig:trilinearsmams}. However, the dependence on $\nicefrac{\kappa_{\Sr}^{(1)}}{\kappa_{\Sr}^{(0)}}$ is
rather weak and inverted due to the additional contributions from quartic couplings which
enhances the dimensionality of the parameter space for $m_s$ and the
contribution of $\kappa_{\Sr}$ to $m_s^{(0)}$ enters with a different sign, cf. 
$m_{22}$ in \cref{eq:mhh,eq:split:mawithoutmatch}. In addition, the
chosen parameter ranges in \cref{tab:split:scanrange} are not optimised for
cases where $m_s<m_h$ which have a relatively small parameter space. Thus, we
find only rarely points with very light singlets (at tree- as well as one-loop
level) such that the maximal loop corrections are smaller than for
$m_a$.
\par
It was already discussed that higher-order corrections, that are bigger than
the leading-order ones, are not necessarily a sign for the breakdown of
perturbation theory, but could be caused 
by accidental cancellations at tree-level. In order to check if this is always
the case, or if there are strongly coupled parameter points, the most reliable
approach is to study the behaviour of the two-loop 
corrections \cite{Braathen:2019pxr}. This is well beyond the scope of this letter, though. Instead, we
use a simpler approach and check the tree-level unitarity constraints using
\SPheno \cite{Goodsell:2018tti} which are
closely connected to perturbativity as discussed in Refs.~\cite{Krauss:2017xpj,Staub:2018vux}.
However, only very few points did not pass the tree-level unitarity
constraints. Thus, perturbativity is presumably not in danger but would require
further investigations.
\par
\begin{figure}[t]
    \centering
    \includegraphics[width=0.42\textwidth]{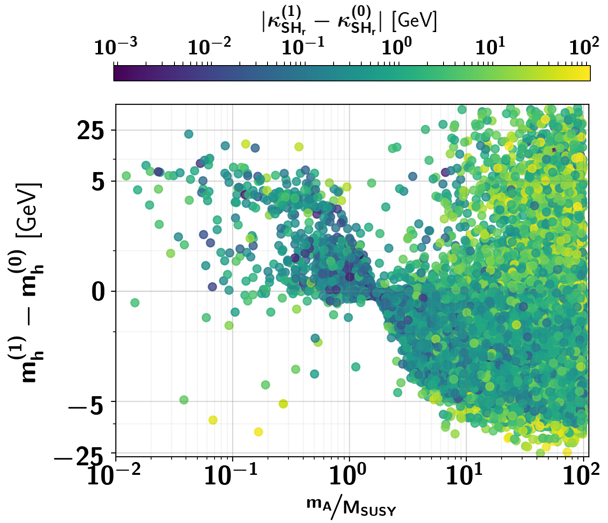}
    \caption{Absolute difference in the prediction of the Higgs boson mass
    using LO/NLO matching conditions as a function of the ratio of the two
    mass scales that are characteristic for sfermions and heavy Higgs bosons.
    All shown points fulfill the Higgs boson mass constraint within a
    \unit[2]{GeV} interval.}
    \label{fig:mhlonlo}
\end{figure}
\begin{figure}[t]
    \centering
    \includegraphics[width=0.4\linewidth]{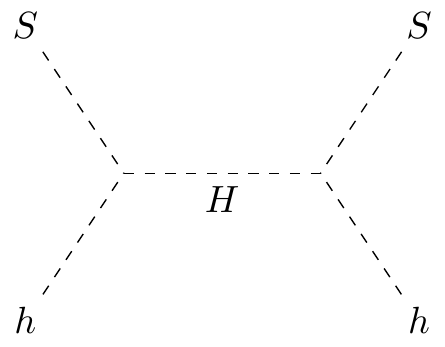}
    \hspace{5mm}
    \includegraphics[width=0.4\linewidth]{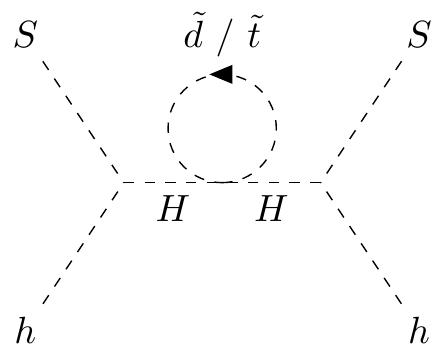}
    \caption{Left diagram: Example of a non-local contribution to the tree-level matching of $\lambda_{SH}$. 
    Right: Example of a $\tan\beta$-enhanced/suppressed one-loop diagram involving a sbottom/stop loop and two heavy Higgs boson propagators.}
        \label{fig:oneloopstops}
\end{figure}

We close the discussion of higher-order corrections with the SM-like Higgs
boson mass corrections shown in \cref{fig:mhlonlo}. In this plot we depict the
absolute Higgs boson mass correction as a function of the 
ratio of the heavy-Higgs mass scale $m_A$ and the matching scale $\msusy$ and
the absolute difference in the coupling $\kappa_{SH_r}$ at $\vev$ when using a
tree-level or a one-loop matching.
The ratio $\nicefrac{m_A}{\msusy}$ appears logarithmically in the matching conditions and becomes crucial 
for the uncertainty of the higher-order corrections for large values. We
observe a negative (positive) shift on the Higgs boson mass for larger (smaller) ratios. Varying
the two scales around 4 orders of magnitude leads to changes in the Higgs
boson mass corrections of more than $\unit[\pm25]{GeV}$. Thus, a tower of EFTs
using e.g. an intermediate matching to a singlet extended 2HDM would be more
reliable if $\nicefrac{m_A}{\msusy}\ll 1$. In addition to a logarithmic
enhancement, diagrams containing a heavy-Higgs propagator with a
down-type sfermion loop attached, can yield and enhancement by positive powers of $\tan\beta$, see for
instance the right diagram in \cref{fig:oneloopstops}. Although $\kappa_{SH_r}$
receives similar contributions like $\kappa_{S_r}$ in \cref{eq:kappa1l}, the
dependence of $m_h$ on $\kappa_{{SH}_r}$ is not as strong as the dependence of $m_{a,s}$ to
$\kappa_{S_r}$. This, is due to the fact, that $\kappa_{SH_r}$ already depends
on $T_\lambda$ at tree-level such that the relative size of
$\kappa_{{SH}_r}^{(0)}$ and $\kappa_{{SH}_r}^{(1)}$ is rather small.
Nevertheless, the impact on the Higgs boson mass 
\begin{equation}
m_h^2\propto v_S \, \kappa_{SH_r}^{(1)},
\end{equation}
can be sizeable for large values of $T_\lambda$ and $\vs$.\\
\par
Despite the logarithmic enhancement, we assume that a one-scale matching is still precise
enough for the considered spread between the two
scales. To answer this question more precisely, a comparison with the two
possible EFT towers is required which is beyond the scope of this letter.

\section{Matching VS. Simplified Models}
\label{sec:results}
\cref{tab:parametercounting} counts the same number of free parameters
contributing to the light scalar masses in the NMSSM as well as the SSM.
Thus, one may ask about the advantage of the matching since the dimensionality
of the parameter space is not reduced. However, it will be shown that SUSY
relations are still active, even if the matching scale is much higher than the electroweak scale. 
This leads to significant constraints on certain parameter regions in the EFT.
\par
A parameter scan according to \cref{tab:split:scanrange} was performed using
the high-scale version of \SPheno as described in \cref{sec:matching}. In addition to the checks performed with \SPheno,
i.e. $\rho$-parameter and tree-level unitarity, we use \micro (version 5.0.2)
\cite{Belanger:2018mqt} to calculate the relic density. Since \micro
is based on the tree-level tool {\tt CalcHEP} \cite{Pukhov:2004ca}, the theoretical uncertainty of
 $\Omega h^2$ is quite high \cite{Bergeron:2017rdm}. Thus, we require $\Omega h^2=0.12\pm
 0.08$. Furthermore, we use \higgsb (version 4.3.1) \cite{Bechtle:2008jh,Bechtle:2011sb,Bechtle:2013wla}, 
 which compares the computed masses, cross-sections and branching fractions against publicly available Higgs searches. In addition to
the Higgs boson mass constraints we also require the lightest chargino to be
heavier than \unit[94]{GeV} and a gluino heavier than \unit[1.5]{TeV}.
\par
\SPheno also offers the possibility to be executed with low-scale input only,
i.e. to perform all low-energy calculations at fixed order using direct inputs
for all Lagrangian parameters in \cref{eq:split:lino,eq:split:left:vgauge}.
For the comparison, a second random scan is performed using this low-scale mode
of \SPheno which is equivalent to common simplified-model approaches. 
In order to produce a comparable parameter set for the SSM, the scan ranges for the low-energy
parameters are chosen from the minimum/maximum values that were the outcome of
the high-scale scan. It should be noted that this procedure is already much more restrictive 
than usual parameter scans of simplified models, were typically all coupling-values allowed by perturbative unitarity 
are in included. The unitarity constraints in the limit of a large scattering energy $s$ and 
negligible differences between the couplings of the CP-even and -odd singlet are 
\begin{align}
& 8 \pi > \text{max}\Big\{|\lambda_H|, |\lambda_S|, |\lambda_{SH}|,   \\
& \frac12 |3 \lambda_H + 2 \lambda_S \pm \sqrt{
   9 \lambda_H^2 + 8 \lambda_{SH}^2 - 12 \lambda_H \lambda_S + 4
   \lambda_S^2}|\Big\}. \nonumber 
\end{align}
If only one of these coupling is large, this corresponds to 
\begin{eqnarray}
& |\lambda^{\text{uni}}_H| < \frac83 \pi \simeq 8.4, \, \, |\lambda^{\text{uni}}_S| < 4 \pi \simeq 12.6, & \nonumber \\
& \text{and}\,\,\,\, |\lambda^{\text{uni}}_{SH}| < 4 \sqrt{2} \pi \simeq 17.8,  &
\end{eqnarray}
which needs to be compared with the range 
\begin{equation}
|\lambda^{\text{EFT}}_H| \lesssim 1.6, \,\,\, |\lambda^{\text{EFT}}_S|
\lesssim 5, \,\,\, |\lambda^{\text{EFT}}_{SH}| \lesssim 3 \;, 
\end{equation}
predicted by the matching. Thus, a conventional simplified-model approach would
have to investigate a parameter space with a volume which is roughly 80 times
larger than those of an NMSSM inspired EFT scan (concerning the subspace of
these three couplings only). Moreover, in the following we show
that this ratio is actually even larger.
\par
\begin{figure}[t]
    \centering
    \includegraphics[width=0.49\textwidth]{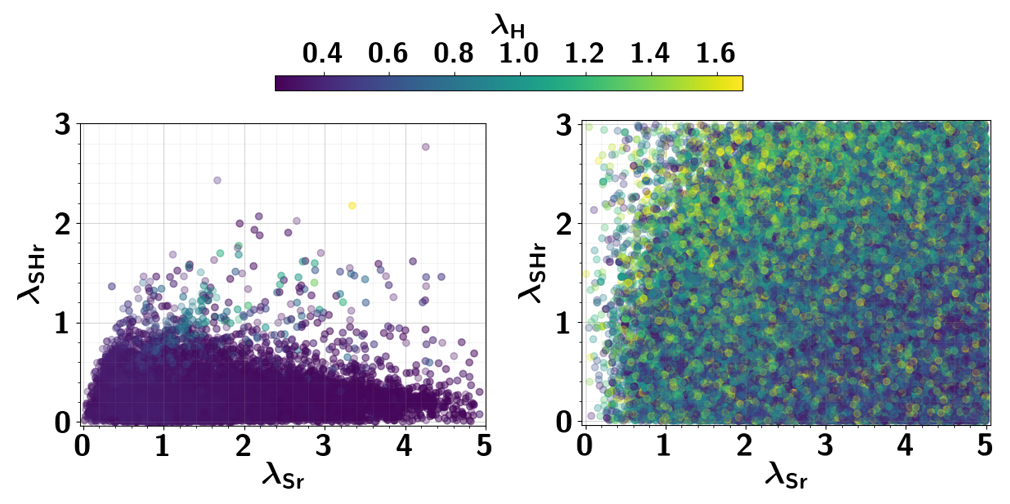}
    \caption{Quartic couplings of the matched EFT (left) and the simplified
    model (right) both evaluated at the scale $\vev$.}
    \label{fig:matching:eftvsuvlam}
\end{figure}

In \cref{fig:matching:eftvsuvlam} we show the correlation between the three
quartic couplings $\lambda_H$, $\lambda_{\Sr}$, $\lambda_{{SH}r}$ evaluated at the
scale $\vev$. 
If a proper matching is performed (left plot), we observe that $\lambda_H$ is small since it is mainly given by $D$-term
contributions. $F$-terms could in principle lead to an enhancement in $\lambda_H$
but would also enhance $\lambda_{{SH}_r}\propto\lambda(\lambda-\kappa)$. Thus,
the singlet-doublet admixture would increase which is already constrained by LHC measurements, i.e. such points 
do not pass the \higgsb checks.
The singlet self-coupling $\lambda_{\Sr}$ is hardly constrained because it affects
only the BSM part of the scalar sector as well as the singlino mass for which no constraint exists.
\par
In comparison, the right plot in \cref{fig:matching:eftvsuvlam} gives the results for the low-scale scan. 
We observe only very weak correlations amongst the three quartic couplings. 
In this case, the collider constraints are the only limiting factor, 
since the scalar couplings are assumed to be independent of each other as well
as of the fermion sector. Thus, a very large fraction of the parameter space
forbidden by SUSY is opened. For very large values of $\lambda_{{SH}_r}$ we
observe a correlation to $\lambda_H$ which implies large cancellations
in the Higgs boson mass in order to achieve a \unit[125]{GeV} Higgs
boson.

\par
\begin{figure}[t]
    \centering
    \includegraphics[width=0.23\textwidth]{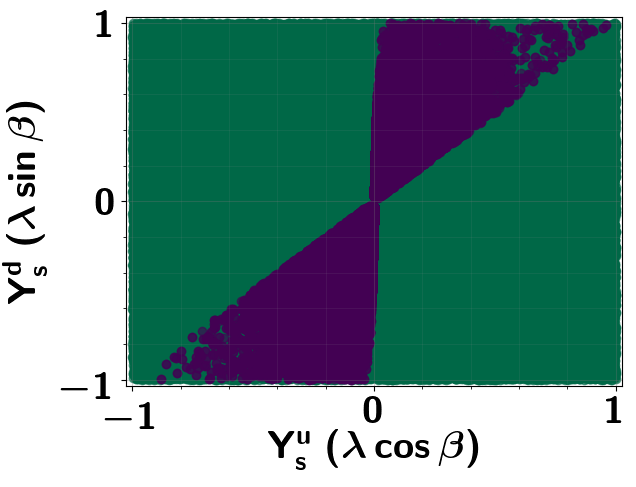}
    \includegraphics[width=0.23\textwidth]{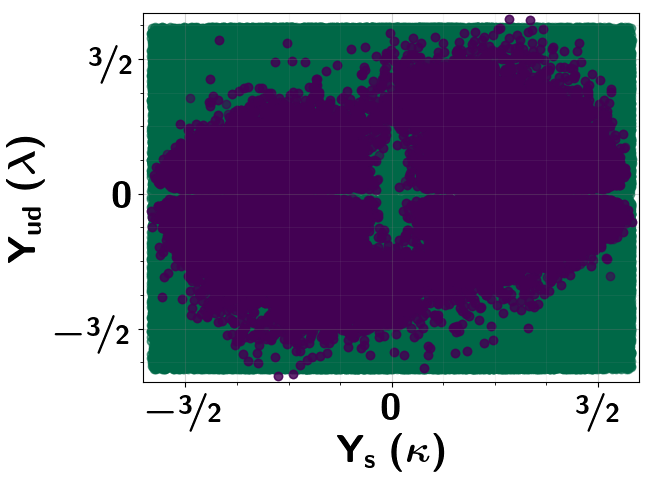}\\
    \includegraphics[width=0.23\textwidth]{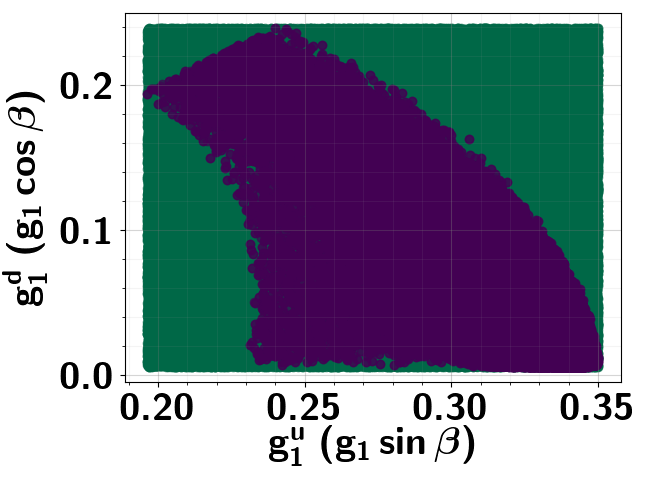}
    \includegraphics[width=0.23\textwidth]{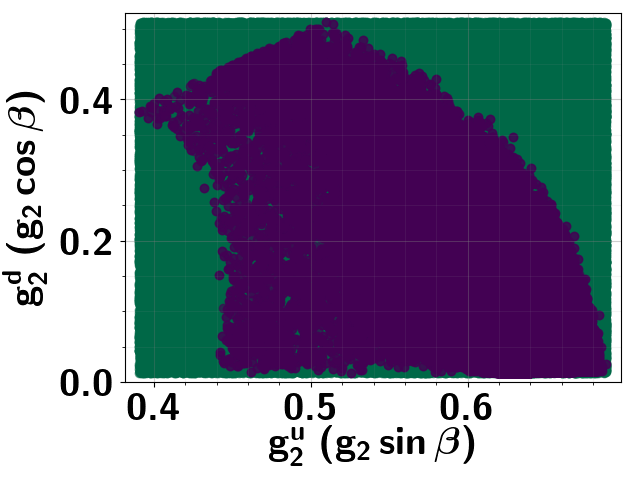}
    \caption{Yukawa couplings of the matched EFT (purple) and the simplified
    model (green) both evaluated at the scale $\vev$. The expressions in
    parentheses are the corresponding tree-level  matching conditions as they
    are applied at the matching scale.}
    \label{fig:matching:eftvsuvyuk}
\end{figure}

The comparison between the two approaches can also be extended to the fermion
sector. In \cref{fig:matching:eftvsuvyuk} we compare all relevant Yukawa
couplings determined with and without the matching to the NMSSM.
Also in this case the EFT approach is actually only compared 
with a small fraction of the parameter space of the simplified model because we are much
more restrictive than perturbative unitarity.
We observe that the chargino constraint is not enough to 
overcome the large freedom in this sector such that the Yukawa couplings in the
simplified model are completely unconstrained. However, during the
matching, SUSY properties give not only relations among different fermionic couplings, 
but also transmit many experimental and theoretical constraints
from the Higgs- to the Yukawa sector. Consequently, a significant fraction of the parameter space is not accessible anymore.
\par
From the considerations on the Lagrangian parameters one can go one step further and
ask the question whether it would be possible to experimentally distinguish
between a SUSY inspired and a pure singlet extension. This question was already
answered positively in Ref.~\cite{Costa:2015llh} based on an analysis of the signal strength of
Higgs-to-Higgs decays in the CxSM\footnote{This is a complex-singlet
extension of the SM with an additional $\mathcal{Z}_2$-symmetry.}
compared to the NMSSM. However, the analysis of the
SSM collider phenomenology is beyond the scope of this letter.
Furthermore, Ref.~\cite{Borschensky:2018zmq} compared DM production in the
MSSM with a simplified model. Also in this comparison, the latter
could not reproduce all phenomenological properties of the
supersymmetric UV completion.
\par
\begin{figure}[t]
    \centering
    \includegraphics[width=0.45\textwidth]{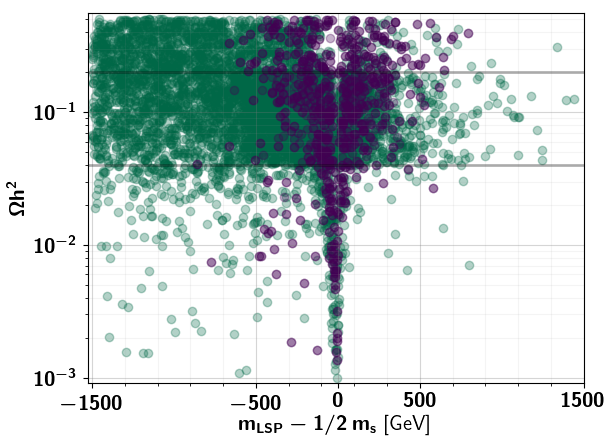}\\
    \caption{Prediction of the relic density using \micro as a
    function of the difference between the singlet-like LSP mass and the
    CP-even scalar singlet in the simplified model (green) and the matched
    EFT (purple). The horizontal lines enclose the region $0.04<\Omega h^2<0.2$ onto
    which we cut in \cref{fig:matching:eftvsuvyuk,fig:matching:eftvsuvlam}. In this
    plot we require at least 90\% singlet admixture in the LSP.}
    \label{fig:matching:omega}
\end{figure}

In contrast to simplified models, SUSY models give hints whether additional states are likely to be found because their mass is close to the one of the
scalar singlet. For instance, when requiring the lightest supersymmetric particle
(LSP) to be a pure singlino, its mass will correlate with the singlet CP-even
mass. This scenario is attractive because it provides an additional mechanism
to avoid a relic density overabundance through resonant annihilation of two DM
singlinos into a singlet-like scalar boson compared to the annihilation mechanism
in the MSSM. In \cref{fig:matching:omega} we
compare this mechanism in both approaches with and without matching. 
We plot the relic density as a function of the mass splitting
$\Delta=m_\text{LSP}-\nicefrac{1}{2} m_s$ while requiring a singlino fraction for the LSP of at least 90\%. 
Consequently $\Omega h^2$ steeply drops near
$\Delta=0$ in both approaches. However, one can also see, that $\Delta$ can
be $\order{\text{TeV}}$ in the simplified model and still being consistent with the dark matter observation.
In contrast, the matching predicts
the two particles to appear within a range of at most \unit[500]{GeV} in order to get a viable dark matter scenario.
\par
Furthermore, if the LSP is Higgsino-like, one can achieve a resonant
annihilation via an s-channel exchange of a CP-odd scalar (with
appropriate mixing). However, this channel is only mildly constrained by
the matching conditions because the CP-odd singlet mass is determined
by the trilinear soft-breaking couplings while the Higgsino masses
scale with $\propto \vs \lambda$.

\section{Conclusion and Outlook}
\label{sec:conclusion}
Simplified models are a widely used method to study specific features of BSM physics.
It is often assumed that these models can arise as the low-energy limit of one or even 
several fundamental UV theories. However, we pointed out that the choice of a concrete UV model can
lead to tremendous constraints on the parameter space of simplified models. 
We have shown for the specific example of the singlet extended SM 
embedded in a SUSY framework, that the full theory always predicts additional light particles to
be present. Thus, simplified models do not have the same power in
giving advise for further experimental searches as the full theory has. Moreover, 
even if all light states are included in the EFT, the predictions between a simplified study
assuming only the parameters at the weak scale, and a full analysis including the proper matching
to the UV theory can be very different. We have shown for several examples that the relations between 
the parameters originating in the UV theory still give large constraints on the accessible parameter space.
This is even the case when there is a large separation between the weak scale and the scale where the 
other degrees of freedom of the full theory are located.
Thus, one needs to make a case by case decision if the usage of simplified models for a specific study is 
justified, or if the matching to the full theory should be included.
\par
If the matching between the EFT and UV theory is considered, the calculation must be done  with the necessary 
precision. We discussed the impact of higher-order corrections of
matching conditions onto the scalar mass spectrum in our chosen setup. We have shown that the effects in the 
singlet sector can be very large under specific conditions. This is in particular the case when there is 
a accidental cancellation at tree-level. In these cases the prediction for the pseudo-scalar mass can 
change by one order of magnitude. As we have discussed, this is {\it not} a
sign of the breakdown of perturbation theory.
\par
This work might deal as guideline for future studies: we have concentrated here on one specific example for the UV and EFT. 
However, also other low-energy limits of the (Split)-NMSSM are possible, e.g.
real singlet extended SM, the 2HDM complemented with a real or 
complex singlet, or just a 2HDM. In addition, contributions from
higher-dimensional operators were neglected, based on experiences from
high-scale MSSM. To verify these results for the split NMSSM, the impact of 
operators like e.g. $S^4 H H^\dagger$ and $S^3 H H^\dagger$ onto the SM-like Higgs
boson mass needs to be investigated.


\section*{Acknowledgements}
We thank Mark Goodsell for discussions about higher-order corrections in scalar matching conditions.
MG acknowledges financal support by the GRK 1694 "Elementary Particle Physics at Highest Energy and highest Precision.
FS is supported by the ERC Recognition Award ERC-RA-0008 of the Helmholtz Association.
This research was supported by the Deutsche Forschungsgemeinschaft (DFG, German Research Foundation) under grant  396021762 - TRR 257.

\bibliography{lit}

\begin{thebibliography}{51}%
\makeatletter
\providecommand \@ifxundefined [1]{%
 \@ifx{#1\undefined}
}%
\providecommand \@ifnum [1]{%
 \ifnum #1\expandafter \@firstoftwo
 \else \expandafter \@secondoftwo
 \fi
}%
\providecommand \@ifx [1]{%
 \ifx #1\expandafter \@firstoftwo
 \else \expandafter \@secondoftwo
 \fi
}%
\providecommand \natexlab [1]{#1}%
\providecommand \enquote  [1]{``#1''}%
\providecommand \bibnamefont  [1]{#1}%
\providecommand \bibfnamefont [1]{#1}%
\providecommand \citenamefont [1]{#1}%
\providecommand \href@noop [0]{\@secondoftwo}%
\providecommand \href [0]{\begingroup \@sanitize@url \@href}%
\providecommand \@href[1]{\@@startlink{#1}\@@href}%
\providecommand \@@href[1]{\endgroup#1\@@endlink}%
\providecommand \@sanitize@url [0]{\catcode `\\12\catcode `\$12\catcode
  `\&12\catcode `\#12\catcode `\^12\catcode `\_12\catcode `\%12\relax}%
\providecommand \@@startlink[1]{}%
\providecommand \@@endlink[0]{}%
\providecommand \url  [0]{\begingroup\@sanitize@url \@url }%
\providecommand \@url [1]{\endgroup\@href {#1}{\urlprefix }}%
\providecommand \urlprefix  [0]{URL }%
\providecommand \Eprint [0]{\href }%
\providecommand \doibase [0]{http://dx.doi.org/}%
\providecommand \selectlanguage [0]{\@gobble}%
\providecommand \bibinfo  [0]{\@secondoftwo}%
\providecommand \bibfield  [0]{\@secondoftwo}%
\providecommand \translation [1]{[#1]}%
\providecommand \BibitemOpen [0]{}%
\providecommand \bibitemStop [0]{}%
\providecommand \bibitemNoStop [0]{.\EOS\space}%
\providecommand \EOS [0]{\spacefactor3000\relax}%
\providecommand \BibitemShut  [1]{\csname bibitem#1\endcsname}%
\let\auto@bib@innerbib\@empty
\bibitem [{\citenamefont {Chatrchyan}\ \emph {et~al.}(2012)\citenamefont
  {Chatrchyan} \emph {et~al.}}]{Chatrchyan:2012ufa}%
  \BibitemOpen
  \bibfield  {author} {\bibinfo {author} {\bibfnamefont {S.}~\bibnamefont
  {Chatrchyan}} \emph {et~al.} (\bibinfo {collaboration} {CMS Collaboration}),\
  }\href {\doibase 10.1016/j.physletb.2012.08.021} {\bibfield  {journal}
  {\bibinfo  {journal} {Phys.Lett.}\ }\textbf {\bibinfo {volume} {B716}},\
  \bibinfo {pages} {30} (\bibinfo {year} {2012})},\ \Eprint
  {http://arxiv.org/abs/1207.7235} {arXiv:1207.7235 [hep-ex]} \BibitemShut
  {NoStop}%
\bibitem [{\citenamefont {Aad}\ \emph {et~al.}(2012)\citenamefont {Aad} \emph
  {et~al.}}]{Aad:2012tfa}%
  \BibitemOpen
  \bibfield  {author} {\bibinfo {author} {\bibfnamefont {G.}~\bibnamefont
  {Aad}} \emph {et~al.} (\bibinfo {collaboration} {ATLAS}),\ }\href {\doibase
  10.1016/j.physletb.2012.08.020} {\bibfield  {journal} {\bibinfo  {journal}
  {Phys. Lett.}\ }\textbf {\bibinfo {volume} {B716}},\ \bibinfo {pages} {1}
  (\bibinfo {year} {2012})},\ \Eprint {http://arxiv.org/abs/1207.7214}
  {arXiv:1207.7214 [hep-ex]} \BibitemShut {NoStop}%
\bibitem [{\citenamefont {Aad}\ \emph {et~al.}(2015)\citenamefont {Aad} \emph
  {et~al.}}]{Aad:2015zhl}%
  \BibitemOpen
  \bibfield  {author} {\bibinfo {author} {\bibfnamefont {G.}~\bibnamefont
  {Aad}} \emph {et~al.} (\bibinfo {collaboration} {ATLAS, CMS}),\ }\href
  {\doibase 10.1103/PhysRevLett.114.191803} {\bibfield  {journal} {\bibinfo
  {journal} {Phys. Rev. Lett.}\ }\textbf {\bibinfo {volume} {114}},\ \bibinfo
  {pages} {191803} (\bibinfo {year} {2015})},\ \Eprint
  {http://arxiv.org/abs/1503.07589} {arXiv:1503.07589 [hep-ex]} \BibitemShut
  {NoStop}%
\bibitem [{\citenamefont {Sirunyan}\ \emph {et~al.}(2018)\citenamefont
  {Sirunyan} \emph {et~al.}}]{Sirunyan:2018vjp}%
  \BibitemOpen
  \bibfield  {author} {\bibinfo {author} {\bibfnamefont {A.~M.}\ \bibnamefont
  {Sirunyan}} \emph {et~al.} (\bibinfo {collaboration} {CMS}),\ }\href
  {\doibase 10.1007/JHEP05(2018)025} {\bibfield  {journal} {\bibinfo  {journal}
  {JHEP}\ }\textbf {\bibinfo {volume} {05}},\ \bibinfo {pages} {025} (\bibinfo
  {year} {2018})},\ \Eprint {http://arxiv.org/abs/1802.02110} {arXiv:1802.02110
  [hep-ex]} \BibitemShut {NoStop}%
\bibitem [{\citenamefont {Wells}(2003)}]{Wells:2003tf}%
  \BibitemOpen
  \bibfield  {author} {\bibinfo {author} {\bibfnamefont {J.~D.}\ \bibnamefont
  {Wells}},\ }in\ \href@noop {} {\emph {\bibinfo {booktitle} {{11th
  International Conference on Supersymmetry and the Unification of Fundamental
  Interactions (SUSY 2003) Tucson, Arizona, June 5-10, 2003}}}}\ (\bibinfo
  {year} {2003})\ \Eprint {http://arxiv.org/abs/hep-ph/0306127}
  {arXiv:hep-ph/0306127 [hep-ph]} \BibitemShut {NoStop}%
\bibitem [{\citenamefont {Arkani-Hamed}\ \emph {et~al.}(2005)\citenamefont
  {Arkani-Hamed}, \citenamefont {Dimopoulos}, \citenamefont {Giudice},\ and\
  \citenamefont {Romanino}}]{ArkaniHamed:2004yi}%
  \BibitemOpen
  \bibfield  {author} {\bibinfo {author} {\bibfnamefont {N.}~\bibnamefont
  {Arkani-Hamed}}, \bibinfo {author} {\bibfnamefont {S.}~\bibnamefont
  {Dimopoulos}}, \bibinfo {author} {\bibfnamefont {G.~F.}\ \bibnamefont
  {Giudice}}, \ and\ \bibinfo {author} {\bibfnamefont {A.}~\bibnamefont
  {Romanino}},\ }\href {\doibase 10.1016/j.nuclphysb.2004.12.026} {\bibfield
  {journal} {\bibinfo  {journal} {Nucl. Phys.}\ }\textbf {\bibinfo {volume}
  {B709}},\ \bibinfo {pages} {3} (\bibinfo {year} {2005})},\ \Eprint
  {http://arxiv.org/abs/hep-ph/0409232} {arXiv:hep-ph/0409232 [hep-ph]}
  \BibitemShut {NoStop}%
\bibitem [{\citenamefont {Arkani-Hamed}\ and\ \citenamefont
  {Dimopoulos}(2005)}]{ArkaniHamed:2004fb}%
  \BibitemOpen
  \bibfield  {author} {\bibinfo {author} {\bibfnamefont {N.}~\bibnamefont
  {Arkani-Hamed}}\ and\ \bibinfo {author} {\bibfnamefont {S.}~\bibnamefont
  {Dimopoulos}},\ }\href {\doibase 10.1088/1126-6708/2005/06/073} {\bibfield
  {journal} {\bibinfo  {journal} {JHEP}\ }\textbf {\bibinfo {volume} {06}},\
  \bibinfo {pages} {073} (\bibinfo {year} {2005})},\ \Eprint
  {http://arxiv.org/abs/hep-th/0405159} {arXiv:hep-th/0405159 [hep-th]}
  \BibitemShut {NoStop}%
\bibitem [{\citenamefont {Giudice}\ and\ \citenamefont
  {Romanino}(2004)}]{Giudice:2004tc}%
  \BibitemOpen
  \bibfield  {author} {\bibinfo {author} {\bibfnamefont {G.~F.}\ \bibnamefont
  {Giudice}}\ and\ \bibinfo {author} {\bibfnamefont {A.}~\bibnamefont
  {Romanino}},\ }\href {\doibase 10.1016/j.nuclphysb.2004.11.048,
  10.1016/j.nuclphysb.2004.08.001} {\bibfield  {journal} {\bibinfo  {journal}
  {Nucl. Phys.}\ }\textbf {\bibinfo {volume} {B699}},\ \bibinfo {pages} {65}
  (\bibinfo {year} {2004})},\ \bibinfo {note} {[Erratum: Nucl.
  Phys.B706,487(2005)]},\ \Eprint {http://arxiv.org/abs/hep-ph/0406088}
  {arXiv:hep-ph/0406088 [hep-ph]} \BibitemShut {NoStop}%
\bibitem [{\citenamefont {Demidov}\ and\ \citenamefont
  {Gorbunov}(2007)}]{Demidov:2006zz}%
  \BibitemOpen
  \bibfield  {author} {\bibinfo {author} {\bibfnamefont {S.~V.}\ \bibnamefont
  {Demidov}}\ and\ \bibinfo {author} {\bibfnamefont {D.~S.}\ \bibnamefont
  {Gorbunov}},\ }\href {\doibase 10.1088/1126-6708/2007/02/055} {\bibfield
  {journal} {\bibinfo  {journal} {JHEP}\ }\textbf {\bibinfo {volume} {02}},\
  \bibinfo {pages} {055} (\bibinfo {year} {2007})},\ \Eprint
  {http://arxiv.org/abs/hep-ph/0612368} {arXiv:hep-ph/0612368 [hep-ph]}
  \BibitemShut {NoStop}%
\bibitem [{\citenamefont {Demidov}\ \emph {et~al.}(2016)\citenamefont
  {Demidov}, \citenamefont {Gorbunov},\ and\ \citenamefont
  {Kirpichnikov}}]{Demidov:2016wcv}%
  \BibitemOpen
  \bibfield  {author} {\bibinfo {author} {\bibfnamefont {S.~V.}\ \bibnamefont
  {Demidov}}, \bibinfo {author} {\bibfnamefont {D.~S.}\ \bibnamefont
  {Gorbunov}}, \ and\ \bibinfo {author} {\bibfnamefont {D.~V.}\ \bibnamefont
  {Kirpichnikov}},\ }\href {\doibase 10.1007/JHEP11(2016)148,
  10.1007/JHEP08(2017)080} {\bibfield  {journal} {\bibinfo  {journal} {JHEP}\
  }\textbf {\bibinfo {volume} {11}},\ \bibinfo {pages} {148} (\bibinfo {year}
  {2016})},\ \bibinfo {note} {[Erratum: JHEP08,080(2017)]},\ \Eprint
  {http://arxiv.org/abs/1608.01985} {arXiv:1608.01985 [hep-ph]} \BibitemShut
  {NoStop}%
\bibitem [{\citenamefont {Demidov}\ \emph {et~al.}(2018)\citenamefont
  {Demidov}, \citenamefont {Gorbunov},\ and\ \citenamefont
  {Kirpichnikov}}]{Demidov:2017lzf}%
  \BibitemOpen
  \bibfield  {author} {\bibinfo {author} {\bibfnamefont {S.~V.}\ \bibnamefont
  {Demidov}}, \bibinfo {author} {\bibfnamefont {D.~S.}\ \bibnamefont
  {Gorbunov}}, \ and\ \bibinfo {author} {\bibfnamefont {D.~V.}\ \bibnamefont
  {Kirpichnikov}},\ }\href {\doibase 10.1016/j.physletb.2018.02.007} {\bibfield
   {journal} {\bibinfo  {journal} {Phys. Lett.}\ }\textbf {\bibinfo {volume}
  {B779}},\ \bibinfo {pages} {191} (\bibinfo {year} {2018})},\ \Eprint
  {http://arxiv.org/abs/1712.00087} {arXiv:1712.00087 [hep-ph]} \BibitemShut
  {NoStop}%
\bibitem [{\citenamefont {Pardo~Vega}\ and\ \citenamefont
  {Villadoro}(2015)}]{Vega:2015fna}%
  \BibitemOpen
  \bibfield  {author} {\bibinfo {author} {\bibfnamefont {J.}~\bibnamefont
  {Pardo~Vega}}\ and\ \bibinfo {author} {\bibfnamefont {G.}~\bibnamefont
  {Villadoro}},\ }\href {\doibase 10.1007/JHEP07(2015)159} {\bibfield
  {journal} {\bibinfo  {journal} {JHEP}\ }\textbf {\bibinfo {volume} {07}},\
  \bibinfo {pages} {159} (\bibinfo {year} {2015})},\ \Eprint
  {http://arxiv.org/abs/1504.05200} {arXiv:1504.05200 [hep-ph]} \BibitemShut
  {NoStop}%
\bibitem [{\citenamefont {Athron}\ \emph {et~al.}(2017)\citenamefont {Athron},
  \citenamefont {Park}, \citenamefont {Steudtner}, \citenamefont
  {Stöckinger},\ and\ \citenamefont {Voigt}}]{Athron:2016fuq}%
  \BibitemOpen
  \bibfield  {author} {\bibinfo {author} {\bibfnamefont {P.}~\bibnamefont
  {Athron}}, \bibinfo {author} {\bibfnamefont {J.-h.}\ \bibnamefont {Park}},
  \bibinfo {author} {\bibfnamefont {T.}~\bibnamefont {Steudtner}}, \bibinfo
  {author} {\bibfnamefont {D.}~\bibnamefont {Stöckinger}}, \ and\ \bibinfo
  {author} {\bibfnamefont {A.}~\bibnamefont {Voigt}},\ }\href {\doibase
  10.1007/JHEP01(2017)079} {\bibfield  {journal} {\bibinfo  {journal} {JHEP}\
  }\textbf {\bibinfo {volume} {01}},\ \bibinfo {pages} {079} (\bibinfo {year}
  {2017})},\ \Eprint {http://arxiv.org/abs/1609.00371} {arXiv:1609.00371
  [hep-ph]} \BibitemShut {NoStop}%
\bibitem [{\citenamefont {Bagnaschi}\ \emph {et~al.}(2017)\citenamefont
  {Bagnaschi}, \citenamefont {Pardo~Vega},\ and\ \citenamefont
  {Slavich}}]{Bagnaschi:2017xid}%
  \BibitemOpen
  \bibfield  {author} {\bibinfo {author} {\bibfnamefont {E.}~\bibnamefont
  {Bagnaschi}}, \bibinfo {author} {\bibfnamefont {J.}~\bibnamefont
  {Pardo~Vega}}, \ and\ \bibinfo {author} {\bibfnamefont {P.}~\bibnamefont
  {Slavich}},\ }\href {\doibase 10.1140/epjc/s10052-017-4885-7} {\bibfield
  {journal} {\bibinfo  {journal} {Eur. Phys. J.}\ }\textbf {\bibinfo {volume}
  {C77}},\ \bibinfo {pages} {334} (\bibinfo {year} {2017})},\ \Eprint
  {http://arxiv.org/abs/1703.08166} {arXiv:1703.08166 [hep-ph]} \BibitemShut
  {NoStop}%
\bibitem [{\citenamefont {Staub}\ and\ \citenamefont
  {Porod}(2017)}]{Staub:2017jnp}%
  \BibitemOpen
  \bibfield  {author} {\bibinfo {author} {\bibfnamefont {F.}~\bibnamefont
  {Staub}}\ and\ \bibinfo {author} {\bibfnamefont {W.}~\bibnamefont {Porod}},\
  }\href {\doibase 10.1140/epjc/s10052-017-4893-7} {\bibfield  {journal}
  {\bibinfo  {journal} {Eur. Phys. J.}\ }\textbf {\bibinfo {volume} {C77}},\
  \bibinfo {pages} {338} (\bibinfo {year} {2017})},\ \Eprint
  {http://arxiv.org/abs/1703.03267} {arXiv:1703.03267 [hep-ph]} \BibitemShut
  {NoStop}%
\bibitem [{\citenamefont {Allanach}\ and\ \citenamefont
  {Voigt}(2018)}]{Allanach:2018fif}%
  \BibitemOpen
  \bibfield  {author} {\bibinfo {author} {\bibfnamefont {B.~C.}\ \bibnamefont
  {Allanach}}\ and\ \bibinfo {author} {\bibfnamefont {A.}~\bibnamefont
  {Voigt}},\ }\href@noop {} {\  (\bibinfo {year} {2018})},\ \Eprint
  {http://arxiv.org/abs/1804.09410} {arXiv:1804.09410 [hep-ph]} \BibitemShut
  {NoStop}%
\bibitem [{\citenamefont {Haber}\ and\ \citenamefont
  {Hempfling}(1993)}]{Haber:1993an}%
  \BibitemOpen
  \bibfield  {author} {\bibinfo {author} {\bibfnamefont {H.~E.}\ \bibnamefont
  {Haber}}\ and\ \bibinfo {author} {\bibfnamefont {R.}~\bibnamefont
  {Hempfling}},\ }\href {\doibase 10.1103/PhysRevD.48.4280} {\bibfield
  {journal} {\bibinfo  {journal} {Phys. Rev.}\ }\textbf {\bibinfo {volume}
  {D48}},\ \bibinfo {pages} {4280} (\bibinfo {year} {1993})},\ \Eprint
  {http://arxiv.org/abs/hep-ph/9307201} {arXiv:hep-ph/9307201 [hep-ph]}
  \BibitemShut {NoStop}%
\bibitem [{\citenamefont {Beneke}\ \emph {et~al.}(2009)\citenamefont {Beneke},
  \citenamefont {Ruiz-Femenia},\ and\ \citenamefont
  {Spinrath}}]{Beneke:2008wj}%
  \BibitemOpen
  \bibfield  {author} {\bibinfo {author} {\bibfnamefont {M.}~\bibnamefont
  {Beneke}}, \bibinfo {author} {\bibfnamefont {P.}~\bibnamefont
  {Ruiz-Femenia}}, \ and\ \bibinfo {author} {\bibfnamefont {M.}~\bibnamefont
  {Spinrath}},\ }\href {\doibase 10.1088/1126-6708/2009/01/031} {\bibfield
  {journal} {\bibinfo  {journal} {JHEP}\ }\textbf {\bibinfo {volume} {01}},\
  \bibinfo {pages} {031} (\bibinfo {year} {2009})},\ \Eprint
  {http://arxiv.org/abs/0810.3768} {arXiv:0810.3768 [hep-ph]} \BibitemShut
  {NoStop}%
\bibitem [{\citenamefont {Gorbahn}\ \emph {et~al.}(2011)\citenamefont
  {Gorbahn}, \citenamefont {Jager}, \citenamefont {Nierste},\ and\
  \citenamefont {Trine}}]{Gorbahn:2009pp}%
  \BibitemOpen
  \bibfield  {author} {\bibinfo {author} {\bibfnamefont {M.}~\bibnamefont
  {Gorbahn}}, \bibinfo {author} {\bibfnamefont {S.}~\bibnamefont {Jager}},
  \bibinfo {author} {\bibfnamefont {U.}~\bibnamefont {Nierste}}, \ and\
  \bibinfo {author} {\bibfnamefont {S.}~\bibnamefont {Trine}},\ }\href
  {\doibase 10.1103/PhysRevD.84.034030} {\bibfield  {journal} {\bibinfo
  {journal} {Phys. Rev.}\ }\textbf {\bibinfo {volume} {D84}},\ \bibinfo {pages}
  {034030} (\bibinfo {year} {2011})},\ \Eprint {http://arxiv.org/abs/0901.2065}
  {arXiv:0901.2065 [hep-ph]} \BibitemShut {NoStop}%
\bibitem [{\citenamefont {Lee}\ and\ \citenamefont
  {Wagner}(2015)}]{Lee:2015uza}%
  \BibitemOpen
  \bibfield  {author} {\bibinfo {author} {\bibfnamefont {G.}~\bibnamefont
  {Lee}}\ and\ \bibinfo {author} {\bibfnamefont {C.~E.~M.}\ \bibnamefont
  {Wagner}},\ }\href {\doibase 10.1103/PhysRevD.92.075032} {\bibfield
  {journal} {\bibinfo  {journal} {Phys. Rev.}\ }\textbf {\bibinfo {volume}
  {D92}},\ \bibinfo {pages} {075032} (\bibinfo {year} {2015})},\ \Eprint
  {http://arxiv.org/abs/1508.00576} {arXiv:1508.00576 [hep-ph]} \BibitemShut
  {NoStop}%
\bibitem [{\citenamefont {Bahl}\ and\ \citenamefont
  {Hollik}(2018)}]{Bahl:2018jom}%
  \BibitemOpen
  \bibfield  {author} {\bibinfo {author} {\bibfnamefont {H.}~\bibnamefont
  {Bahl}}\ and\ \bibinfo {author} {\bibfnamefont {W.}~\bibnamefont {Hollik}},\
  }\href@noop {} {\  (\bibinfo {year} {2018})},\ \Eprint
  {http://arxiv.org/abs/1805.00867} {arXiv:1805.00867 [hep-ph]} \BibitemShut
  {NoStop}%
\bibitem [{\citenamefont {Gabelmann}\ \emph {et~al.}(2019)\citenamefont
  {Gabelmann}, \citenamefont {Muehlleitner},\ and\ \citenamefont
  {Staub}}]{Gabelmann:2018axh}%
  \BibitemOpen
  \bibfield  {author} {\bibinfo {author} {\bibfnamefont {M.}~\bibnamefont
  {Gabelmann}}, \bibinfo {author} {\bibfnamefont {M.}~\bibnamefont
  {Muehlleitner}}, \ and\ \bibinfo {author} {\bibfnamefont {F.}~\bibnamefont
  {Staub}},\ }\href {\doibase 10.1140/epjc/s10052-019-6570-5} {\bibfield
  {journal} {\bibinfo  {journal} {Eur. Phys. J.}\ }\textbf {\bibinfo {volume}
  {C79}},\ \bibinfo {pages} {163} (\bibinfo {year} {2019})},\ \Eprint
  {http://arxiv.org/abs/1810.12326} {arXiv:1810.12326 [hep-ph]} \BibitemShut
  {NoStop}%
\bibitem [{\citenamefont {Bahl}(2019)}]{Bahl:2019ccs}%
  \BibitemOpen
  \bibfield  {author} {\bibinfo {author} {\bibfnamefont {H.}~\bibnamefont
  {Bahl}},\ }in\ \href@noop {} {\emph {\bibinfo {booktitle} {{54th Rencontres
  de Moriond on QCD and High Energy Interactions (Moriond QCD 2019) La Thuile,
  Italy, March 23-30, 2019}}}}\ (\bibinfo {year} {2019})\ \Eprint
  {http://arxiv.org/abs/1905.04918} {arXiv:1905.04918 [hep-ph]} \BibitemShut
  {NoStop}%
\bibitem [{\citenamefont {Alves}(2012)}]{Alves:2011wf}%
  \BibitemOpen
  \bibfield  {author} {\bibinfo {author} {\bibfnamefont {D.}~\bibnamefont
  {Alves}} (\bibinfo {collaboration} {LHC New Physics Working Group}),\ }\href
  {\doibase 10.1088/0954-3899/39/10/105005} {\bibfield  {journal} {\bibinfo
  {journal} {J. Phys.}\ }\textbf {\bibinfo {volume} {G39}},\ \bibinfo {pages}
  {105005} (\bibinfo {year} {2012})},\ \Eprint {http://arxiv.org/abs/1105.2838}
  {arXiv:1105.2838 [hep-ph]} \BibitemShut {NoStop}%
\bibitem [{\citenamefont {Contino}\ \emph {et~al.}(2013)\citenamefont
  {Contino}, \citenamefont {Ghezzi}, \citenamefont {Grojean}, \citenamefont
  {Muhlleitner},\ and\ \citenamefont {Spira}}]{Contino:2013kra}%
  \BibitemOpen
  \bibfield  {author} {\bibinfo {author} {\bibfnamefont {R.}~\bibnamefont
  {Contino}}, \bibinfo {author} {\bibfnamefont {M.}~\bibnamefont {Ghezzi}},
  \bibinfo {author} {\bibfnamefont {C.}~\bibnamefont {Grojean}}, \bibinfo
  {author} {\bibfnamefont {M.}~\bibnamefont {Muhlleitner}}, \ and\ \bibinfo
  {author} {\bibfnamefont {M.}~\bibnamefont {Spira}},\ }\href {\doibase
  10.1007/JHEP07(2013)035} {\bibfield  {journal} {\bibinfo  {journal} {JHEP}\
  }\textbf {\bibinfo {volume} {07}},\ \bibinfo {pages} {035} (\bibinfo {year}
  {2013})},\ \Eprint {http://arxiv.org/abs/1303.3876} {arXiv:1303.3876
  [hep-ph]} \BibitemShut {NoStop}%
\bibitem [{\citenamefont {Costa}\ \emph {et~al.}(2015)\citenamefont {Costa},
  \citenamefont {Morais}, \citenamefont {Sampaio},\ and\ \citenamefont
  {Santos}}]{Costa:2014qga}%
  \BibitemOpen
  \bibfield  {author} {\bibinfo {author} {\bibfnamefont {R.}~\bibnamefont
  {Costa}}, \bibinfo {author} {\bibfnamefont {A.~P.}\ \bibnamefont {Morais}},
  \bibinfo {author} {\bibfnamefont {M.~O.~P.}\ \bibnamefont {Sampaio}}, \ and\
  \bibinfo {author} {\bibfnamefont {R.}~\bibnamefont {Santos}},\ }\href
  {\doibase 10.1103/PhysRevD.92.025024} {\bibfield  {journal} {\bibinfo
  {journal} {Phys. Rev.}\ }\textbf {\bibinfo {volume} {D92}},\ \bibinfo {pages}
  {025024} (\bibinfo {year} {2015})},\ \Eprint {http://arxiv.org/abs/1411.4048}
  {arXiv:1411.4048 [hep-ph]} \BibitemShut {NoStop}%
\bibitem [{\citenamefont {Buttazzo}\ \emph {et~al.}(2015)\citenamefont
  {Buttazzo}, \citenamefont {Sala},\ and\ \citenamefont
  {Tesi}}]{Buttazzo:2015bka}%
  \BibitemOpen
  \bibfield  {author} {\bibinfo {author} {\bibfnamefont {D.}~\bibnamefont
  {Buttazzo}}, \bibinfo {author} {\bibfnamefont {F.}~\bibnamefont {Sala}}, \
  and\ \bibinfo {author} {\bibfnamefont {A.}~\bibnamefont {Tesi}},\ }\href
  {\doibase 10.1007/JHEP11(2015)158} {\bibfield  {journal} {\bibinfo  {journal}
  {JHEP}\ }\textbf {\bibinfo {volume} {11}},\ \bibinfo {pages} {158} (\bibinfo
  {year} {2015})},\ \Eprint {http://arxiv.org/abs/1505.05488} {arXiv:1505.05488
  [hep-ph]} \BibitemShut {NoStop}%
\bibitem [{\citenamefont {Costa}\ \emph {et~al.}(2016)\citenamefont {Costa},
  \citenamefont {Mühlleitner}, \citenamefont {Sampaio},\ and\ \citenamefont
  {Santos}}]{Costa:2015llh}%
  \BibitemOpen
  \bibfield  {author} {\bibinfo {author} {\bibfnamefont {R.}~\bibnamefont
  {Costa}}, \bibinfo {author} {\bibfnamefont {M.}~\bibnamefont {Mühlleitner}},
  \bibinfo {author} {\bibfnamefont {M.~O.~P.}\ \bibnamefont {Sampaio}}, \ and\
  \bibinfo {author} {\bibfnamefont {R.}~\bibnamefont {Santos}},\ }\href
  {\doibase 10.1007/JHEP06(2016)034} {\bibfield  {journal} {\bibinfo  {journal}
  {JHEP}\ }\textbf {\bibinfo {volume} {06}},\ \bibinfo {pages} {034} (\bibinfo
  {year} {2016})},\ \Eprint {http://arxiv.org/abs/1512.05355} {arXiv:1512.05355
  [hep-ph]} \BibitemShut {NoStop}%
\bibitem [{\citenamefont {Borschensky}\ \emph {et~al.}(2019)\citenamefont
  {Borschensky}, \citenamefont {Coniglio},\ and\ \citenamefont
  {Jäger}}]{Borschensky:2018zmq}%
  \BibitemOpen
  \bibfield  {author} {\bibinfo {author} {\bibfnamefont {C.}~\bibnamefont
  {Borschensky}}, \bibinfo {author} {\bibfnamefont {G.}~\bibnamefont
  {Coniglio}}, \ and\ \bibinfo {author} {\bibfnamefont {B.}~\bibnamefont
  {Jäger}},\ }\href {\doibase 10.1140/epjc/s10052-019-6945-7} {\bibfield
  {journal} {\bibinfo  {journal} {Eur. Phys. J.}\ }\textbf {\bibinfo {volume}
  {C79}},\ \bibinfo {pages} {428} (\bibinfo {year} {2019})},\ \Eprint
  {http://arxiv.org/abs/1812.08704} {arXiv:1812.08704 [hep-ph]} \BibitemShut
  {NoStop}%
\bibitem [{\citenamefont {Martin}(1997)}]{Martin:1997ns}%
  \BibitemOpen
  \bibfield  {author} {\bibinfo {author} {\bibfnamefont {S.~P.}\ \bibnamefont
  {Martin}},\ }\href {\doibase 10.1142/9789812839657_0001,
  10.1142/9789814307505_0001} {\ ,\ \bibinfo {pages} {1} (\bibinfo {year}
  {1997})},\ \bibinfo {note} {[Adv. Ser. Direct. High Energy
  Phys.18,1(1998)]},\ \Eprint {http://arxiv.org/abs/hep-ph/9709356}
  {arXiv:hep-ph/9709356 [hep-ph]} \BibitemShut {NoStop}%
\bibitem [{\citenamefont {Maniatis}(2010)}]{Maniatis:2009re}%
  \BibitemOpen
  \bibfield  {author} {\bibinfo {author} {\bibfnamefont {M.}~\bibnamefont
  {Maniatis}},\ }\href {\doibase 10.1142/S0217751X10049827} {\bibfield
  {journal} {\bibinfo  {journal} {Int.J.Mod.Phys.}\ }\textbf {\bibinfo {volume}
  {A25}},\ \bibinfo {pages} {3505} (\bibinfo {year} {2010})},\ \Eprint
  {http://arxiv.org/abs/0906.0777} {arXiv:0906.0777 [hep-ph]} \BibitemShut
  {NoStop}%
\bibitem [{\citenamefont {Ellwanger}\ \emph {et~al.}(2010)\citenamefont
  {Ellwanger}, \citenamefont {Hugonie},\ and\ \citenamefont
  {Teixeira}}]{Ellwanger:2009dp}%
  \BibitemOpen
  \bibfield  {author} {\bibinfo {author} {\bibfnamefont {U.}~\bibnamefont
  {Ellwanger}}, \bibinfo {author} {\bibfnamefont {C.}~\bibnamefont {Hugonie}},
  \ and\ \bibinfo {author} {\bibfnamefont {A.~M.}\ \bibnamefont {Teixeira}},\
  }\href {\doibase 10.1016/j.physrep.2010.07.001} {\bibfield  {journal}
  {\bibinfo  {journal} {Phys. Rept.}\ }\textbf {\bibinfo {volume} {496}},\
  \bibinfo {pages} {1} (\bibinfo {year} {2010})},\ \Eprint
  {http://arxiv.org/abs/0910.1785} {arXiv:0910.1785} \BibitemShut {NoStop}%
\bibitem [{\citenamefont {Goodsell}\ \emph {et~al.}(2015)\citenamefont
  {Goodsell}, \citenamefont {Nickel},\ and\ \citenamefont
  {Staub}}]{Goodsell:2014pla}%
  \BibitemOpen
  \bibfield  {author} {\bibinfo {author} {\bibfnamefont {M.~D.}\ \bibnamefont
  {Goodsell}}, \bibinfo {author} {\bibfnamefont {K.}~\bibnamefont {Nickel}}, \
  and\ \bibinfo {author} {\bibfnamefont {F.}~\bibnamefont {Staub}},\ }\href
  {\doibase 10.1103/PhysRevD.91.035021} {\bibfield  {journal} {\bibinfo
  {journal} {Phys. Rev.}\ }\textbf {\bibinfo {volume} {D91}},\ \bibinfo {pages}
  {035021} (\bibinfo {year} {2015})},\ \Eprint {http://arxiv.org/abs/1411.4665}
  {arXiv:1411.4665 [hep-ph]} \BibitemShut {NoStop}%
\bibitem [{\citenamefont {Bagnaschi}\ \emph {et~al.}(2014)\citenamefont
  {Bagnaschi}, \citenamefont {Giudice}, \citenamefont {Slavich},\ and\
  \citenamefont {Strumia}}]{Bagnaschi:2014rsa}%
  \BibitemOpen
  \bibfield  {author} {\bibinfo {author} {\bibfnamefont {E.}~\bibnamefont
  {Bagnaschi}}, \bibinfo {author} {\bibfnamefont {G.~F.}\ \bibnamefont
  {Giudice}}, \bibinfo {author} {\bibfnamefont {P.}~\bibnamefont {Slavich}}, \
  and\ \bibinfo {author} {\bibfnamefont {A.}~\bibnamefont {Strumia}},\ }\href
  {\doibase 10.1007/JHEP09(2014)092} {\bibfield  {journal} {\bibinfo  {journal}
  {JHEP}\ }\textbf {\bibinfo {volume} {09}},\ \bibinfo {pages} {092} (\bibinfo
  {year} {2014})},\ \Eprint {http://arxiv.org/abs/1407.4081} {arXiv:1407.4081
  [hep-ph]} \BibitemShut {NoStop}%
\bibitem [{\citenamefont {Staub}(2010)}]{Staub:2009bi}%
  \BibitemOpen
  \bibfield  {author} {\bibinfo {author} {\bibfnamefont {F.}~\bibnamefont
  {Staub}},\ }\href {\doibase 10.1016/j.cpc.2010.01.011} {\bibfield  {journal}
  {\bibinfo  {journal} {Comput.Phys.Commun.}\ }\textbf {\bibinfo {volume}
  {181}},\ \bibinfo {pages} {1077} (\bibinfo {year} {2010})},\ \Eprint
  {http://arxiv.org/abs/0909.2863} {arXiv:0909.2863 [hep-ph]} \BibitemShut
  {NoStop}%
\bibitem [{\citenamefont {Staub}(2011)}]{Staub:2010jh}%
  \BibitemOpen
  \bibfield  {author} {\bibinfo {author} {\bibfnamefont {F.}~\bibnamefont
  {Staub}},\ }\href {\doibase 10.1016/j.cpc.2010.11.030} {\bibfield  {journal}
  {\bibinfo  {journal} {Comput.Phys.Commun.}\ }\textbf {\bibinfo {volume}
  {182}},\ \bibinfo {pages} {808} (\bibinfo {year} {2011})},\ \Eprint
  {http://arxiv.org/abs/1002.0840} {arXiv:1002.0840 [hep-ph]} \BibitemShut
  {NoStop}%
\bibitem [{\citenamefont {Staub}(2013)}]{Staub:2012pb}%
  \BibitemOpen
  \bibfield  {author} {\bibinfo {author} {\bibfnamefont {F.}~\bibnamefont
  {Staub}},\ }\href {\doibase 10.1016/j.cpc.2013.02.019} {\bibfield  {journal}
  {\bibinfo  {journal} {Comput. Phys. Commun.}\ }\textbf {\bibinfo {volume}
  {184}},\ \bibinfo {pages} {1792} (\bibinfo {year} {2013})},\ \Eprint
  {http://arxiv.org/abs/1207.0906} {arXiv:1207.0906 [hep-ph]} \BibitemShut
  {NoStop}%
\bibitem [{\citenamefont {Staub}(2014)}]{Staub:2013tta}%
  \BibitemOpen
  \bibfield  {author} {\bibinfo {author} {\bibfnamefont {F.}~\bibnamefont
  {Staub}},\ }\href {\doibase 10.1016/j.cpc.2014.02.018} {\bibfield  {journal}
  {\bibinfo  {journal} {Comput. Phys. Commun.}\ }\textbf {\bibinfo {volume}
  {185}},\ \bibinfo {pages} {1773} (\bibinfo {year} {2014})},\ \Eprint
  {http://arxiv.org/abs/1309.7223} {arXiv:1309.7223 [hep-ph]} \BibitemShut
  {NoStop}%
\bibitem [{\citenamefont {Braathen}\ \emph {et~al.}(2018)\citenamefont
  {Braathen}, \citenamefont {Goodsell},\ and\ \citenamefont
  {Slavich}}]{Braathen:2018htl}%
  \BibitemOpen
  \bibfield  {author} {\bibinfo {author} {\bibfnamefont {J.}~\bibnamefont
  {Braathen}}, \bibinfo {author} {\bibfnamefont {M.~D.}\ \bibnamefont
  {Goodsell}}, \ and\ \bibinfo {author} {\bibfnamefont {P.}~\bibnamefont
  {Slavich}},\ }\href@noop {} {\  (\bibinfo {year} {2018})},\ \Eprint
  {http://arxiv.org/abs/1810.09388} {arXiv:1810.09388 [hep-ph]} \BibitemShut
  {NoStop}%
\bibitem [{\citenamefont {Porod}(2003)}]{Porod:2003um}%
  \BibitemOpen
  \bibfield  {author} {\bibinfo {author} {\bibfnamefont {W.}~\bibnamefont
  {Porod}},\ }\href {\doibase 10.1016/S0010-4655(03)00222-4} {\bibfield
  {journal} {\bibinfo  {journal} {Comput.Phys.Commun.}\ }\textbf {\bibinfo
  {volume} {153}},\ \bibinfo {pages} {275} (\bibinfo {year} {2003})},\ \Eprint
  {http://arxiv.org/abs/hep-ph/0301101} {arXiv:hep-ph/0301101 [hep-ph]}
  \BibitemShut {NoStop}%
\bibitem [{\citenamefont {Porod}\ and\ \citenamefont
  {Staub}(2011)}]{Porod:2011nf}%
  \BibitemOpen
  \bibfield  {author} {\bibinfo {author} {\bibfnamefont {W.}~\bibnamefont
  {Porod}}\ and\ \bibinfo {author} {\bibfnamefont {F.}~\bibnamefont {Staub}},\
  }\href@noop {} {\  (\bibinfo {year} {2011})},\ \Eprint
  {http://arxiv.org/abs/1104.1573} {arXiv:1104.1573 [hep-ph]} \BibitemShut
  {NoStop}%
\bibitem [{\citenamefont {Braathen}\ and\ \citenamefont
  {Kanemura}(2019)}]{Braathen:2019pxr}%
  \BibitemOpen
  \bibfield  {author} {\bibinfo {author} {\bibfnamefont {J.}~\bibnamefont
  {Braathen}}\ and\ \bibinfo {author} {\bibfnamefont {S.}~\bibnamefont
  {Kanemura}},\ }\href {\doibase 10.1016/j.physletb.2019.07.021} {\bibfield
  {journal} {\bibinfo  {journal} {Phys. Lett.}\ }\textbf {\bibinfo {volume}
  {B796}},\ \bibinfo {pages} {38} (\bibinfo {year} {2019})},\ \Eprint
  {http://arxiv.org/abs/1903.05417} {arXiv:1903.05417 [hep-ph]} \BibitemShut
  {NoStop}%
\bibitem [{\citenamefont {Goodsell}\ and\ \citenamefont
  {Staub}(2018)}]{Goodsell:2018tti}%
  \BibitemOpen
  \bibfield  {author} {\bibinfo {author} {\bibfnamefont {M.~D.}\ \bibnamefont
  {Goodsell}}\ and\ \bibinfo {author} {\bibfnamefont {F.}~\bibnamefont
  {Staub}},\ }\href@noop {} {\  (\bibinfo {year} {2018})},\ \Eprint
  {http://arxiv.org/abs/1805.07306} {arXiv:1805.07306 [hep-ph]} \BibitemShut
  {NoStop}%
\bibitem [{\citenamefont {Krauss}\ and\ \citenamefont
  {Staub}(2018)}]{Krauss:2017xpj}%
  \BibitemOpen
  \bibfield  {author} {\bibinfo {author} {\bibfnamefont {M.~E.}\ \bibnamefont
  {Krauss}}\ and\ \bibinfo {author} {\bibfnamefont {F.}~\bibnamefont {Staub}},\
  }\href {\doibase 10.1140/epjc/s10052-018-5676-5} {\bibfield  {journal}
  {\bibinfo  {journal} {Eur. Phys. J.}\ }\textbf {\bibinfo {volume} {C78}},\
  \bibinfo {pages} {185} (\bibinfo {year} {2018})},\ \Eprint
  {http://arxiv.org/abs/1709.03501} {arXiv:1709.03501 [hep-ph]} \BibitemShut
  {NoStop}%
\bibitem [{\citenamefont {Staub}(2019)}]{Staub:2018vux}%
  \BibitemOpen
  \bibfield  {author} {\bibinfo {author} {\bibfnamefont {F.}~\bibnamefont
  {Staub}},\ }\href {\doibase 10.1016/j.physletb.2018.12.039} {\bibfield
  {journal} {\bibinfo  {journal} {Phys. Lett.}\ }\textbf {\bibinfo {volume}
  {B789}},\ \bibinfo {pages} {203} (\bibinfo {year} {2019})},\ \Eprint
  {http://arxiv.org/abs/1811.08300} {arXiv:1811.08300 [hep-ph]} \BibitemShut
  {NoStop}%
\bibitem [{\citenamefont {Bélanger}\ \emph {et~al.}(2018)\citenamefont
  {Bélanger}, \citenamefont {Boudjema}, \citenamefont {Goudelis},
  \citenamefont {Pukhov},\ and\ \citenamefont {Zaldivar}}]{Belanger:2018mqt}%
  \BibitemOpen
  \bibfield  {author} {\bibinfo {author} {\bibfnamefont {G.}~\bibnamefont
  {Bélanger}}, \bibinfo {author} {\bibfnamefont {F.}~\bibnamefont {Boudjema}},
  \bibinfo {author} {\bibfnamefont {A.}~\bibnamefont {Goudelis}}, \bibinfo
  {author} {\bibfnamefont {A.}~\bibnamefont {Pukhov}}, \ and\ \bibinfo {author}
  {\bibfnamefont {B.}~\bibnamefont {Zaldivar}},\ }\href {\doibase
  10.1016/j.cpc.2018.04.027} {\bibfield  {journal} {\bibinfo  {journal}
  {Comput. Phys. Commun.}\ }\textbf {\bibinfo {volume} {231}},\ \bibinfo
  {pages} {173} (\bibinfo {year} {2018})},\ \Eprint
  {http://arxiv.org/abs/1801.03509} {arXiv:1801.03509 [hep-ph]} \BibitemShut
  {NoStop}%
\bibitem [{\citenamefont {Pukhov}(2004)}]{Pukhov:2004ca}%
  \BibitemOpen
  \bibfield  {author} {\bibinfo {author} {\bibfnamefont {A.}~\bibnamefont
  {Pukhov}},\ }\href@noop {} {\  (\bibinfo {year} {2004})},\ \Eprint
  {http://arxiv.org/abs/hep-ph/0412191} {arXiv:hep-ph/0412191 [hep-ph]}
  \BibitemShut {NoStop}%
\bibitem [{\citenamefont {Bergeron}\ \emph {et~al.}(2018)\citenamefont
  {Bergeron}, \citenamefont {Sandick},\ and\ \citenamefont
  {Sinha}}]{Bergeron:2017rdm}%
  \BibitemOpen
  \bibfield  {author} {\bibinfo {author} {\bibfnamefont {P.}~\bibnamefont
  {Bergeron}}, \bibinfo {author} {\bibfnamefont {P.}~\bibnamefont {Sandick}}, \
  and\ \bibinfo {author} {\bibfnamefont {K.}~\bibnamefont {Sinha}},\ }\href
  {\doibase 10.1007/JHEP05(2018)113} {\bibfield  {journal} {\bibinfo  {journal}
  {JHEP}\ }\textbf {\bibinfo {volume} {05}},\ \bibinfo {pages} {113} (\bibinfo
  {year} {2018})},\ \Eprint {http://arxiv.org/abs/1712.05491} {arXiv:1712.05491
  [hep-ph]} \BibitemShut {NoStop}%
\bibitem [{\citenamefont {Bechtle}\ \emph {et~al.}(2010)\citenamefont
  {Bechtle}, \citenamefont {Brein}, \citenamefont {Heinemeyer}, \citenamefont
  {Weiglein},\ and\ \citenamefont {Williams}}]{Bechtle:2008jh}%
  \BibitemOpen
  \bibfield  {author} {\bibinfo {author} {\bibfnamefont {P.}~\bibnamefont
  {Bechtle}}, \bibinfo {author} {\bibfnamefont {O.}~\bibnamefont {Brein}},
  \bibinfo {author} {\bibfnamefont {S.}~\bibnamefont {Heinemeyer}}, \bibinfo
  {author} {\bibfnamefont {G.}~\bibnamefont {Weiglein}}, \ and\ \bibinfo
  {author} {\bibfnamefont {K.~E.}\ \bibnamefont {Williams}},\ }\href {\doibase
  10.1016/j.cpc.2009.09.003} {\bibfield  {journal} {\bibinfo  {journal}
  {Comput. Phys. Commun.}\ }\textbf {\bibinfo {volume} {181}},\ \bibinfo
  {pages} {138} (\bibinfo {year} {2010})},\ \Eprint
  {http://arxiv.org/abs/0811.4169} {arXiv:0811.4169 [hep-ph]} \BibitemShut
  {NoStop}%
\bibitem [{\citenamefont {Bechtle}\ \emph {et~al.}(2011)\citenamefont
  {Bechtle}, \citenamefont {Brein}, \citenamefont {Heinemeyer}, \citenamefont
  {Weiglein},\ and\ \citenamefont {Williams}}]{Bechtle:2011sb}%
  \BibitemOpen
  \bibfield  {author} {\bibinfo {author} {\bibfnamefont {P.}~\bibnamefont
  {Bechtle}}, \bibinfo {author} {\bibfnamefont {O.}~\bibnamefont {Brein}},
  \bibinfo {author} {\bibfnamefont {S.}~\bibnamefont {Heinemeyer}}, \bibinfo
  {author} {\bibfnamefont {G.}~\bibnamefont {Weiglein}}, \ and\ \bibinfo
  {author} {\bibfnamefont {K.~E.}\ \bibnamefont {Williams}},\ }\href {\doibase
  10.1016/j.cpc.2011.07.015} {\bibfield  {journal} {\bibinfo  {journal}
  {Comput.Phys.Commun.}\ }\textbf {\bibinfo {volume} {182}},\ \bibinfo {pages}
  {2605} (\bibinfo {year} {2011})},\ \Eprint {http://arxiv.org/abs/1102.1898}
  {arXiv:1102.1898 [hep-ph]} \BibitemShut {NoStop}%
\bibitem [{\citenamefont {Bechtle}\ \emph {et~al.}(2014)\citenamefont
  {Bechtle}, \citenamefont {Brein}, \citenamefont {Heinemeyer}, \citenamefont
  {Stal}, \citenamefont {Stefaniak}, \citenamefont {Weiglein},\ and\
  \citenamefont {Williams}}]{Bechtle:2013wla}%
  \BibitemOpen
  \bibfield  {author} {\bibinfo {author} {\bibfnamefont {P.}~\bibnamefont
  {Bechtle}}, \bibinfo {author} {\bibfnamefont {O.}~\bibnamefont {Brein}},
  \bibinfo {author} {\bibfnamefont {S.}~\bibnamefont {Heinemeyer}}, \bibinfo
  {author} {\bibfnamefont {O.}~\bibnamefont {Stal}}, \bibinfo {author}
  {\bibfnamefont {T.}~\bibnamefont {Stefaniak}}, \bibinfo {author}
  {\bibfnamefont {G.}~\bibnamefont {Weiglein}}, \ and\ \bibinfo {author}
  {\bibfnamefont {K.~E.}\ \bibnamefont {Williams}},\ }\href {\doibase
  10.1140/epjc/s10052-013-2693-2} {\bibfield  {journal} {\bibinfo  {journal}
  {Eur. Phys. J.}\ }\textbf {\bibinfo {volume} {C74}},\ \bibinfo {pages} {2693}
  (\bibinfo {year} {2014})},\ \Eprint {http://arxiv.org/abs/1311.0055}
  {arXiv:1311.0055 [hep-ph]} \BibitemShut {NoStop}%
\end{thebibliography}%

\end{document}